\documentclass[reprint,prb]{revtex4-2}
\usepackage[utf8]{inputenc}
\usepackage{amsmath,amsfonts,amssymb}
\usepackage{graphicx}
\graphicspath{{/figures}}
\usepackage{mathtools}
\usepackage{bbold}
\usepackage{bm}
\usepackage[colorlinks,allcolors = blue]{hyperref}
\usepackage[section]{placeins}
\usepackage[capitalise]{cleveref}
\usepackage{color}
\usepackage{siunitx}
\usepackage{nicefrac}
\usepackage{booktabs}

\newcommand{\liou}[1][]{\mathcal{L}_\mathrm{#1}}
\newcommand{\comm}[2]{\left[ #1,#2\right]}
\newcommand{\hamil}[1][]{H_\mathrm{#1}}
\newcommand{\ev}[1]{\langle #1 \rangle}
\newcommand{\evz}[1]{\langle #1 \rangle_0}
\newcommand{\iu}{\mathrm{i}}

\usepackage{adjustbox}\newcommand{\op}[2][]{#2_{#1}}

\newcommand{\ket}[1]{\left|#1\rangle \right.}
\newcommand{\bra}[1]{\langle\left.#1\right| }
\newcommand{\liouket}[1]{\left|#1\rangle\rangle \right.}

\newcommand{\Tr}[2][ ]{\mathrm{Tr}_\mathrm{#1}\left\{#2\right\}}
\newcommand{\gsz}{\bar{\Gamma}^\mathrm{sub}}
\newcommand{\gtz}{\bar{\Gamma}^\mathrm{tip}}
\newcommand{\gez}{\bar{\Gamma}^\eta}
\newcommand{\proj}[1][]{\mathcal{P}^{\mathit{#1}}}

\DeclareMathOperator{\e}{\mathrm{e}}

\newcommand{\fmt}{f^-_\mathrm{tip}(\epsilon_{\mathrm{t}})}
\newcommand{\fpt}{f^+_\mathrm{tip}(\epsilon_{\mathrm{t}})}
\newcommand{\fms}{f^-_\mathrm{sub}(\epsilon_{\mathrm{t}})}

\newcommand{\fmtS}{f^-_\mathrm{tip}(\epsilon_\mathrm{s})}
\newcommand{\fptS}{f^+_\mathrm{tip}(\epsilon_\mathrm{s})}
\newcommand{\fmsS}{f^-_\mathrm{sub}(\epsilon_\mathrm{s})}

\newcommand{\fpe}{f^+_\eta(\epsilon_\mathrm{t})}

\newcommand{\fpeS}{f^+_\eta(\epsilon_\mathrm{s})}

\begin{document}

\title{Spin-orbit interaction induces charge beatings in a 
lightwave-STM --- single molecule junction}
\author{Moritz Frankerl}
\author{Andrea Donarini}
\email{andrea.donarini@ur.de}
\affiliation{Institute for Theoretical Physics, University of Regensburg, 93040 Regensburg, Germany}
\date{\today}
\begin{abstract}
Recent lightwave-STM experiments have shown space and time resolution of single molecule vibrations directly on their intrinsic length and time-scales. We address here theoretically the electronic dynamics of a copper-phthalocyanine in a lightwave-STM, explored within a pump-probe cycle scheme. The spin-orbit interaction in the metallic center induces beatings of the electric charge flowing through the molecule as a function of the delay time between the pump and the probe pulses. Interference between the quasidegenerate anionic states of the molecule and the intertwined dynamics of the associated spin and pseudospin degrees of freedom are the key aspects of such phenomenon. We study the dynamics directly in the time domain within a generalized master-equation approach.
\end{abstract}
\maketitle
%
\section{Introduction}

The spin-orbit interaction (SOI), the relativistic phenomenon which locks the spin and orbital degrees of freedom, is key in the emerging field of molecular spintronics \cite{Sanvito2011}. It improves our capabilities to read, write and control spin states. It is therefore of utmost importance to gain a rigorous understanding and a precise control of the SOI on the level of the microscopic building blocks of condensed matter itself. 

One may consider a gedanken experiment and observe what happens when a single electron is injected into one individual molecule. The electron is subject to the spin-orbit interaction and, at the same time, it exchanges energy with the local environment. Generally, extremely short lifetimes of the electronic spin state are expected. The ideal experiment would hence access electronic dynamics simultaneously on the atomic length scale and on time-scales as short as femtoseconds. Such an ultimate experiment, fully resolving the dynamics of individual spins in space and time, can actually be implemented based on the latest breakthroughs of scanning tunneling microscopy (STM) and ultrax fast photonics \cite{Cocker2016,Peller2020}. We address in this work its theoretical simulation.

Imaging, spectroscopy, and manipulation with atomic resolution has been achieved with steady-state STM. Such experiments have provided unique insights into the equilibrium electronic states of molecules \cite{Joachim1995,Repp2005,Liu2015,Schulz2015,Yu2017,Reecht2020}. Resolving dynamics directly in the time domain, however, requires excitations to be confined to a short time window. An all-electronic pump-probe scheme has been realized in the seminal work by Loth and coworkers \cite{Loth2010}. They achieved a direct observation of the relaxation of individual spins on the atomic length scale and the nanosecond timescale. A time resolution up to a few picoseconds has been progressively reached by optical means \cite{Hamers1990,Nunes1993,Terada2010}. In parallel, ultrafast photonics has started to explore processes on time-scales faster even than a single cycle of light. Thus, it became possible to stir electronic motion directly via the oscillating carrier wave of tailored light pulses – a principle often dubbed “lightwave electronics”. 
\begin{figure}
\begin{center}
    \includegraphics[width=\linewidth]{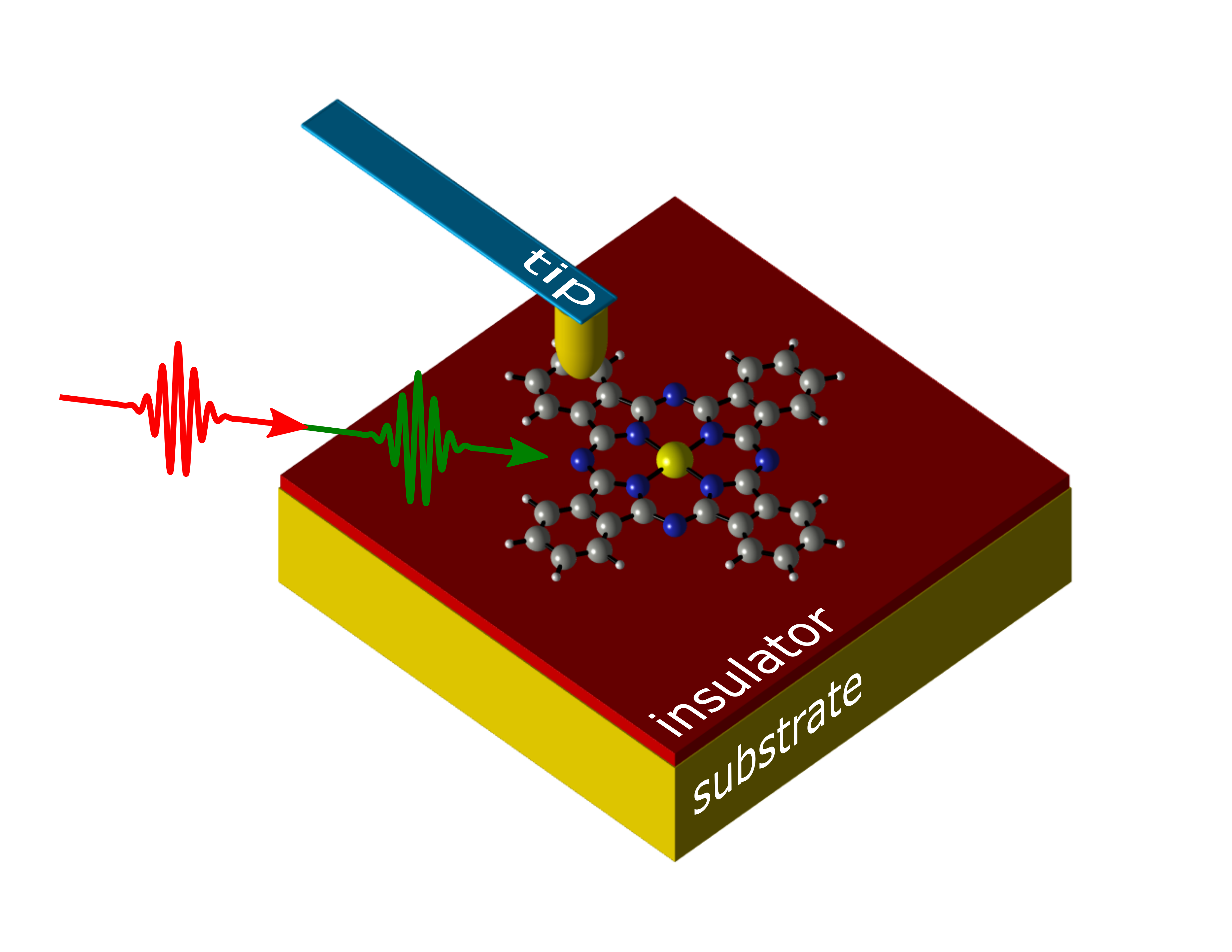}
\end{center}
\caption{Schematic representation of a lightwave-STM single molecule junction. The molecule is adsorbed on an insulator to hinder its hybridization to the underlying metallic substrate. The sharp tip allows one to image the molecule with atomic resolution. The bias across the junction is modulated by laser pulses, illustrated by the green (red) wave packet for the pump (probe) pulse.}
\label{fig:setup}
\end{figure}

Cocker et al.~combine in Ref. \cite{Cocker2016} for the  first time THz lightwave electronics with scanning tunneling microscopy to initiate and monitor the center-of-mass oscillations of an individual molecule on the atomic length and subpicosecond time-scale. THz pulses act in their experiments analogously to the voltage pulses in the all-electronic pump-probe experiments of Ref. \cite{Loth2010}, but they access much shorter timescales.

In our theoretical investigation, we propose a pump probe scheme to initiate and observe the SOI-induced dynamics on a copper phthalocyanine (CuPc) molecule \cref{fig:setup}. 
Differently from our previous works on this molecule \cite{Donarini2012,Siegert2015a,Siegert2016}, in which we considered exclusively the steady state transport and tunneling spectroscopy, we now focus our analysis directly on the time-resolved response of the molecule and investigate the intertwined spin and pseudospin dynamics induced by the spin-orbit coupling on the central copper atom. To this end, we extend our model and dynamical description of the molecular junction to fully simulate the pump probe cycle of a lightwave-STM.

\begin{figure}[h!]
\begin{center}
    \includegraphics[width=\linewidth]{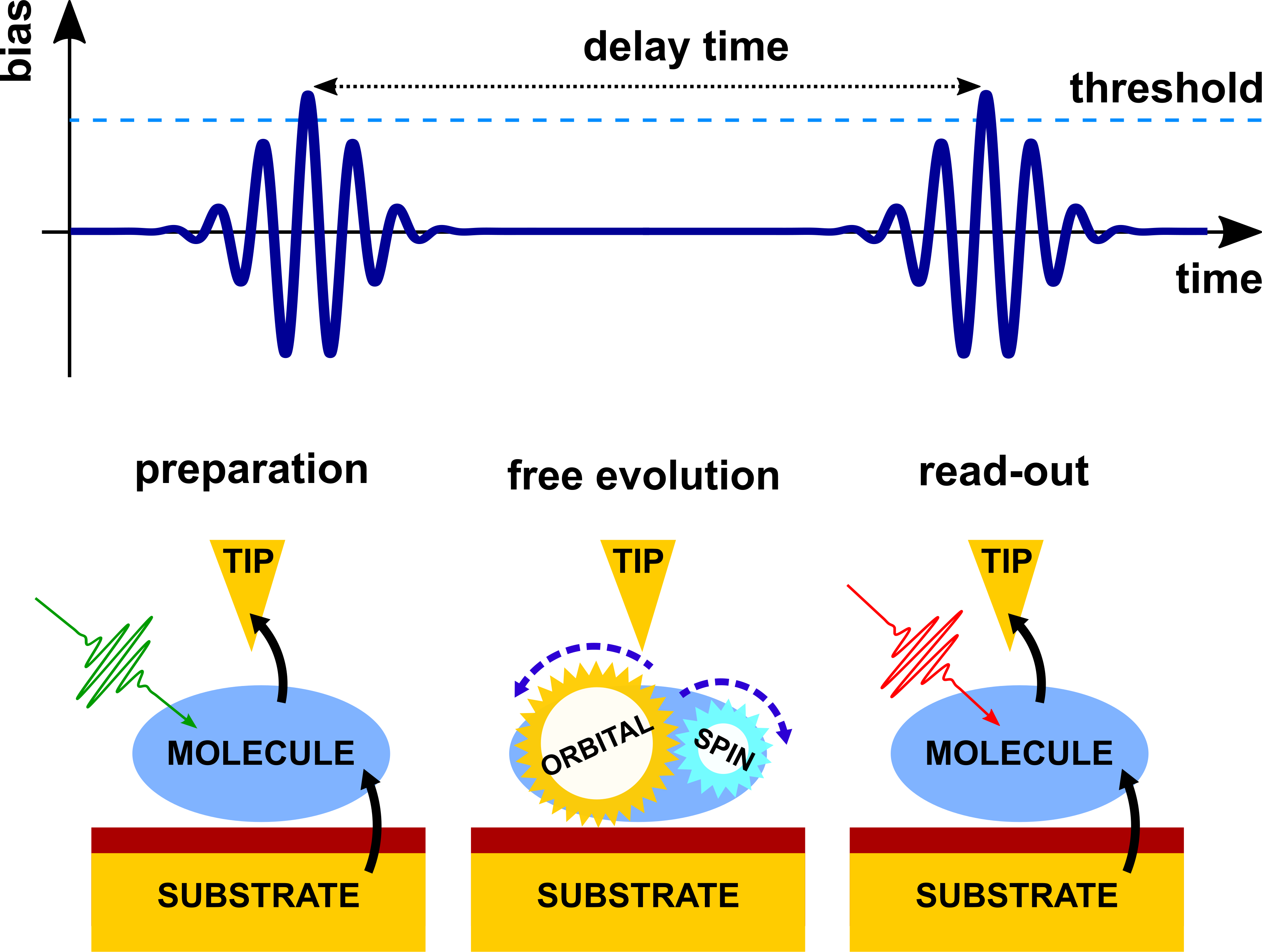}
\end{center}
\caption{Pump-probe scheme in lightwave-STM single molecule junction. The THz pulse induces a bias pulse across the junction. Charge flows whenever this transient bias pulse overcomes a threshold. The first pulse brings CuPc, due to interference, into a superposition of many-body states. The intertwined spin-orbital dynamics on the molecule dominates the free-evolution during the delay time and thus influences the read-out signal, i.e., the transferred charge into the tip averaged over several pump-probe cycles.}
\label{fig:pump_probe}
\end{figure}

As schematically described in \cref{fig:pump_probe} and in close analogy to the experiments reported in Refs. \cite{Cocker2016,Peller2020}, the bias pulses induced by the incoming lightwaves trigger tunneling events between the molecule and the leads. During the delay time between the two pulses, instead, the molecule evolves under equilibrated electrochemical potentials. The SOI strongly permeates this time lapse, yielding an intertwined evolution of the spin and orbital degrees of freedom. The latter is then revealed by a probe pulse as the transferred charge per pump-probe cycle measured as a function of the delay time. 

A crucial role is played in our proposed scheme by the interference effects associated with interacting nanojunctions with quasi orbital degeneracy \cite{Michaelis2006,Donarini2009,Donarini2010,Nilsson2010}, recently observed in suspended CNTs \cite{Donarini2019}. CuPc does exhibit orbital degeneracy, protected by symmetry, and the pump pulse brings the molecule from its thermal equilibrium into a coherent superposition of the quasidegenerate many-body states which blocks the current. The resulting electrical dark state is characterized by a well-defined pseudospin, which, during the delay time, exhibits  precession and beating dynamics under the influence of the SOI. 

The average charge flowing through the STM junction during the probe pulse critically depends on the state reached by the molecule at the beginning of this second  pulse. The closer is the molecule to the dark state, the less current flows. Thus, the SOI-induced dynamics leaves its fingerprints in the average charge  per pump-probe cycle collected at the tip.     

We calculate the transport characteristics of this driven and interacting single-molecule junction directly in the time domain, by means of a generalized master equation for the reduced density matrix of the molecule. Two complementary approaches to driven transport have been taken in the literature: approximated methods adapted to the parameter configuration of specific systems or numerically exact methods. 

Prominent members of the first class are the Keldysh \cite{Jauho1994, Platero2004} and scattering approaches \cite{Platero2004, Kohler2005} to driven systems or quantum master equations \cite{Platero2004,Kohler2005} that are especially suited for molecules with weak coupling to the leads. In the presence of a time-dependent driving field the energy is no longer a conserved quantity, and the evaluation of the time evolution operator becomes a nontrivial task. Simplifications are possible for time-periodic perturbations, allowing the use of Floquet theory \cite{Grifoni1998}. Hence, to reduce complexity, most of the theoretical works so far have considered driven molecular bridges either in a single-particle approximation \cite{Volkovich2011, Donarini2012b, Selzer2013, Galperin2012a} or in extremely simplified interacting models \cite{Platero2004,Lopez1998,Takafumi2015}. 

The second class of methods comprises numerically exact approaches: no perturbation parameter is introduced here, at a price of making these methods numerically extremely demanding. Because of this, electron-electron interactions and driving are considered for very simplified models of the molecular bridge \cite{Nghiem2014,Dou2020}. Recently, a quantum master-equation approach has been derived for a driven interacting molecular bridge, with arbitrary shape of the driving field \cite{Agarwalla2015,Peskin2016}. 

We derive the quantum master equation for our system in the Markov approximation, relying on the fast relaxation of the electronic correlations in the metallic leads. We treat both the coupling to the leads as well as the SOI perturbatively and assume an adiabatic limit with respect to the driving speed. Yet the dynamics of CuPc is nonadiabatic, due to the matching of the time-scales for the tunneling events and the SOI-induced precession. Moreover, we investigate also the influence of the Lamb shift correction to the coherent dynamics of the molecule.  Such a contribution derives from the virtual electronic fluctuations between the molecule and the metallic leads. Due to the orbital degeneracies, it plays a significant role in the time evolution of the system, mainly during the time lapse between the two laser pulses. 

The modeling of the Coulomb interaction and the SOI on the CuPc is clearly of fundamental importance in our transport calculations.  
Spin-orbit coupling is known to be essential, in combination with the configuration of the nonmagnetic component (organic ligand), in establishing the magnetic anisotropy in high-spin molecular magnets \cite{Gatteschi2006}. Effective spin Hamiltonians are commonly used to describe this anisotropy, and usually capture the low-energy properties of these systems well \cite{Gatteschi2006,Mannini2010}. Such effective Hamiltonians have been derived microscopically for the widely studied molecular magnets like Fe$_8$, Fe$_4$ and Mn$_{12}$ \cite{Chiesa2013}. We have explicitly investigated long-range and short-range electron-electron correlation effects in CuPc \cite{Siegert2016}. By adding the SOI to our analysis, we obtained a magnetic anisotropy which can in turn be captured by an effective spin Hamiltonian \cite{Siegert2015a}. 

In general, the accurate calculation of the many-body properties of metal-organic molecules is a highly nontrivial task since the number of atomic constituents is large enough for exact diagonalization to be impossible, while standard density-functional schemes have difficulty capturing short-ranged electron-electron correlations \cite{Chiesa2013}. To reduce the size of the many-body Fock space, we use a basis of frontier molecular orbitals as the starting point to include electronic correlations and construct a generalized Hubbard Hamiltonian \cite{Siegert2016}, an approach similar to the one presented in Ref.~\cite{Chiesa2013}.

The paper is structured as follows: In Sec.~\ref{sec:model} we first introduce the model Hamiltonian for CuPc, with focus on the low-energy spectrum and eigenstates of the neutral and ionic molecule. The full description of the lightwave-STM is given in Sec.~\ref{sec:lightwave-STM}, with particular emphasis on the form of the tunneling matrices. Moreover, in Sec.~\ref{sec:transport} we derive the generalized master equation for the reduced density matrix (RDM) and give the formula for the evaluation of the current in terms of this matrix. The latter is parametrized in Sec.~\ref{sec:Operator_space} by means of the product operator basis for the Liouville space. The intricate dynamics of the molecule during the pump pulse are described in Sec.~\ref{sec:prep}, introducing, step by step, first the tunneling dynamics, then the influence of the SOI, and finally the Lamb-shift contribution to the state preparation. Free evolution and read out are discussed in Sec.~\ref{sec:evolution}. We concentrate on the pseudospin beatings revealing the intertwined dynamics of the spin and orbital degrees of freedom. Finally, we discuss the read-out mechanism implemented by the probe pulse. Concluding remarks and perspectives are given in Sec.~\ref{sec:conclusions}.      

\section{Model and Transport theory}

We introduce first the many-body Hamiltonian of the molecule under investigation and highlight the interplay of exchange and spin-orbit interaction in the characterization of its low-energy eigenstates. 

\label{sec:model}
\subsection{Copper phthalocyanine}

\begin{figure}
    \centering
    \includegraphics[width=0.8\linewidth]{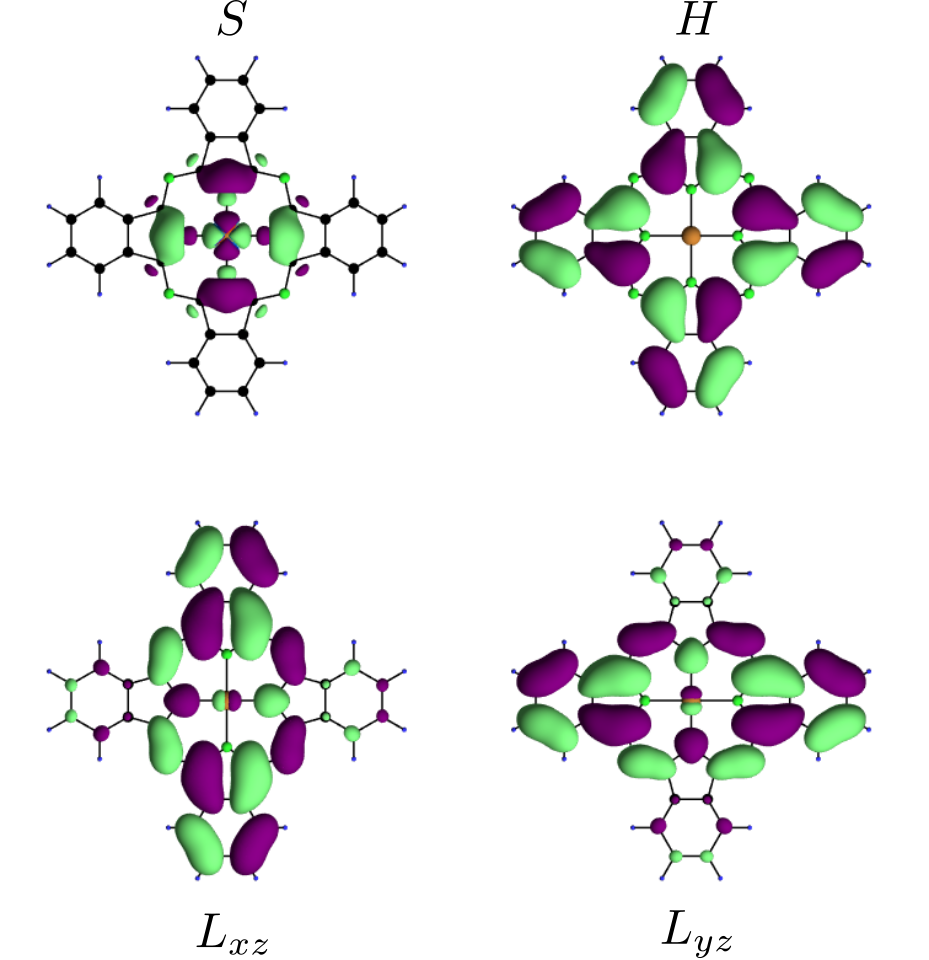}     
    \caption{Isosurfaces of the four frontier orbitals: SOMO, HOMO and, the real-valued LUMOs. The color indicates the sign of the wave function at the isosurface. Green corresponds to a positive,  purple to a negative sign.}
    \label{fig:real_lumos}
\end{figure}

Copper-phthalocyanine (CuPc) is a metal-organic compound with $D_{4h}$ symmetry consisting of an organic ligand surrounding a copper atom (see \cref{fig:setup}). We model the molecule with the Hamiltonian
\begin{equation}
    \label{eq:hamil_molecule}
    \hamil[M] = \hamil[SP] + \op[\rm ee]{V} + \hamil[SOI] = \hamil[0] + \hamil[SOI],
\end{equation}
distinguishing the single-particle $\hamil[SP]$, the electron-electron interaction $\op[\rm ee]{V}$, and the spin-orbit interaction $\hamil[SOI]$ component, respectively. The spin-orbit interaction is treated perturbatively, due to its smaller energy scales as compared with those of $\hamil[0]$. We are interested into the low-energy states with the molecule being at most two times negatively charged (CuPc$^{2-}$). With reference to the interacting model derived by Siegert et. al.~\cite{Siegert2015a,Siegert2016} we project the different terms of Eq.~\eqref{eq:hamil_molecule} on the many-body basis constructed with all possible occupations of four frontier orbitals (see Fig.~\ref{fig:real_lumos}). Listed with increasing single-particle energy, these are, respectively, the singly occupied molecular orbital (SOMO or simply $S$ in the following), the highest occupied molecular orbital (HOMO or $H$) and the two degenerate lowest unoccupied molecular orbitals (LUMOs). The latter are degenerate and thus admit different representations. We have chosen in Fig.~\ref{fig:real_lumos} the one with real valued wave functions labeled $L_{xz}$ and $L_{yz}$) to recall their symmetry, the same of the atomic $d$-shell orbitals  $d_{xz}$, and $d_{yz}$, with respect to the $D_{4h}$ group. 
The degenerate real LUMOs can be brought in their complex, rotationally invariant, form 
\begin{equation}
\label{eq:lumopm}
    \ket{L\pm} =\frac{1}{\sqrt{2}} \big( \ket{L_{xz} }\pm \iu\ket{L_{yz}}\big),
\end{equation}
which will also be used in our discussions.
The single-particle Hamiltonian and the Coulomb interactions have, in the basis of the frontier orbitals, the generic form
\begin{equation}
    \label{eq:hamil_unperturbed}
\hamil[0] = \sum_i \epsilon_i \op[i]{n} + \frac{1}{2}\sum_{ijkl}\sum_{\sigma\sigma'}
    V_{ijkl} \op[i\sigma]{d}^\dag \op[k\sigma']{d}^\dag\op[l\sigma']{d}\op[j\sigma]{d}, 
\end{equation}
where $i$, $j$, $k$, $l=S,H,L+,\,\text{and}\,L-$, while $\sigma$ represents the spin and $\op[i]{n}= \op[\sigma i]{d}^\dag\op[i\sigma]{d}$ is the number operator for the $i$-th molecular orbital.
The single-particle energies obtained from a tight-binding calculation and renormalized by a crystal-field correction \cite{Siegert2015a,Siegert2016} are $\epsilon_S= -\SI{10.17}{\eV}$,
$\epsilon_H = -\SI{9.87}{\eV}$, $\epsilon_{L\pm}=-\SI{8.87}{\eV}$. All Coulomb integrals $V_{ijkl}$ have been included in the numerical evaluation of $\hamil[M]$. Several of those integrals, though, vanish due to symmetry arguments (i.e., invariance of the Hamiltonian $\hamil[0]$ with respect to the $D_{4h}$ point symmetry group).  We retain, in the numerical diagonalization, all the nonvanishing contributions, given in \cref{tab:coulomb}. 
The large difference between the Hubbard energy of the SOMO with respect to those of the HOMO justifies the single occupation of the first in the neutral ground state of the molecule despite the ordering of the single-particle orbital energies, which sees the SOMO below the HOMO level.  

\setlength{\tabcolsep}{8.5pt}
\begin{table}
    \centering
\begin{tabular}[c]{l r  l r}
    \toprule[1.5pt]
    $U_\mathrm{S}$ & 11.352 eV&$J^{\mathrm{ex}}_{\mathrm{HL}}=-\tilde{J}^P_{\mathrm{H+-}}$&548 meV\\
    \addlinespace
    $U_\mathrm{H}$ & 1.752 eV & $J^{\mathrm{ex}}_{+-}$  &  258  meV\\
    \addlinespace
    $U_\mathrm{+/-} = U_{+-}$ & 1.808 eV & $J^P_{+-}$  &  168 meV\\
    \addlinespace
    $U_\mathrm{SH}$ & 1.777 eV & $J^{\mathrm{ex}}_{\mathrm{SL}}=-\tilde{J}^P_{\mathrm{S+-}}$&9 meV\\
    \addlinespace
    $U_\mathrm{SL}$ & 1.993 eV & $J^{\mathrm{ex}}_{\mathrm{SH}} = J^P_{\mathrm{SH}}$  &  2 meV\\
    \addlinespace
    $U_\mathrm{HL}$ & 1.758 eV &   &  \\
    \bottomrule
\end{tabular}
\caption{Coulomb integrals between the frontier orbitals. We indicate with $U_i = V_{iiii}$ the Hubbard energy of the orbital $i$, with $U_{ij} = V_{iijj}$  and $J^{\rm ex}_{ij} = V_{ijji}$ being the direct Coulomb integral and the exchange between the orbitals $i$ and $j$, . Finally, $J^{\rm P}_{ij} = V_{ijij}$ and $\tilde{J}^{\rm P}_{ijk} = V_{ijik}$ refer to the pair hopping and the split pair hopping, respectively.  Taken from Ref.~\cite{Siegert2015a}.}
\label{tab:coulomb}
\end{table}

The SOI Hamiltonian for the molecule is the superposition of the atomic ones 
\begin{equation}
    \hamil[SOI]= \sum_{\alpha} \xi_{\alpha} \op[\alpha]{\vec{l}}\cdot\op[\alpha]{\vec{s}},
\end{equation}
where $\alpha$ is a collective index which runs over all atoms and shells, $\op[\alpha]{\vec{l}}$ is the corresponding angular-momentum operator and $\op[\alpha]{\vec{s}}$ is the spin operator. We concentrate, though, only on the $d$ shell of the copper atom, which carries by far the largest spin-orbit coupling $\xi_\alpha$. Projecting this contribution into the basis of the frontier orbitals yields eventually 
\begin{equation}
\begin{aligned}     
    \hamil[SOI] = &\lambda_1 \sum_{\tau=\pm}\tau \left(\op[L\tau\uparrow]{d}^\dag\op[L\tau\uparrow]{d}-
\op[L\tau\downarrow]{d}^\dag \op[L\tau\downarrow]{d}\right) \\+
    &\lambda_2\left(\op[S\uparrow]{d}^\dag\op[L-\downarrow]{d}+\op[L+\uparrow]{d}^\dag\op[S\downarrow]{d}+ 
    \mathrm{H.c.}  \right) ,
\end{aligned}     
\end{equation}
with the effective SOI parameters $\lambda_1 = \SI{0.47}{\meV}$ and $\lambda_2 = \SI{6.16}{\meV}$. The size of the effective SOI parameters with respect to the bare one of the copper $d$ shell $\xi_{\rm Cu} \approx \SI{100}{meV}$ \cite{Bendix1993} reflects the relatively low weight of the LUMOs on the copper atom (see Fig.~\ref{fig:real_lumos}). 

The electronic transitions induced by the tunneling events between the leads and the molecule involve the many-body eigenstates of the molecule. To retain correlation effects, we calculate them via the exact diagonalization of the Hamiltonian in \cref{eq:hamil_molecule} within the full configuration space of the four frontier orbitals depicted in \cref{fig:real_lumos}. We focus here on the neutral and anionic low-energy states, within an energy range of a few $\SI{}{meV}$. Only these states will play a role in our transport calculations.

The eigenstates of the Hamiltonian in Eq.~\eqref{eq:hamil_molecule} are linear combinations of several Slater determinants which we fully retain in the  numerical calculations. The admixtures are though very small and we neglect them in the analytical description of the eigenstates presented below. To this end, it is useful to define a state $\ket{\Omega} = \op[H\uparrow]{d}^\dag\op[H\downarrow]{d}^\dag\ket{0}$, and to express the energy eigenstates in terms of $\ket{\Omega}$. 

The neutral ground state is a spin degenerate doublet with a predominant contributions of the states
\begin{equation}
\ket{D^\uparrow} = \op[S\uparrow]{d}^\dag\ket{\Omega}, \quad
    \ket{D^\downarrow} = \op[S\downarrow]{d}^\dag\ket{\Omega},
\end{equation}
where we observe the characteristic unpaired spin in the SOMO.  The first-excited neutral energy level lies about $\SI{0.8}{\eV}$ above the ground state \cite{Siegert2016} and thus we do not consider it in our calculations.  

\begin{figure}
    \centering
    \includegraphics[width=\linewidth]{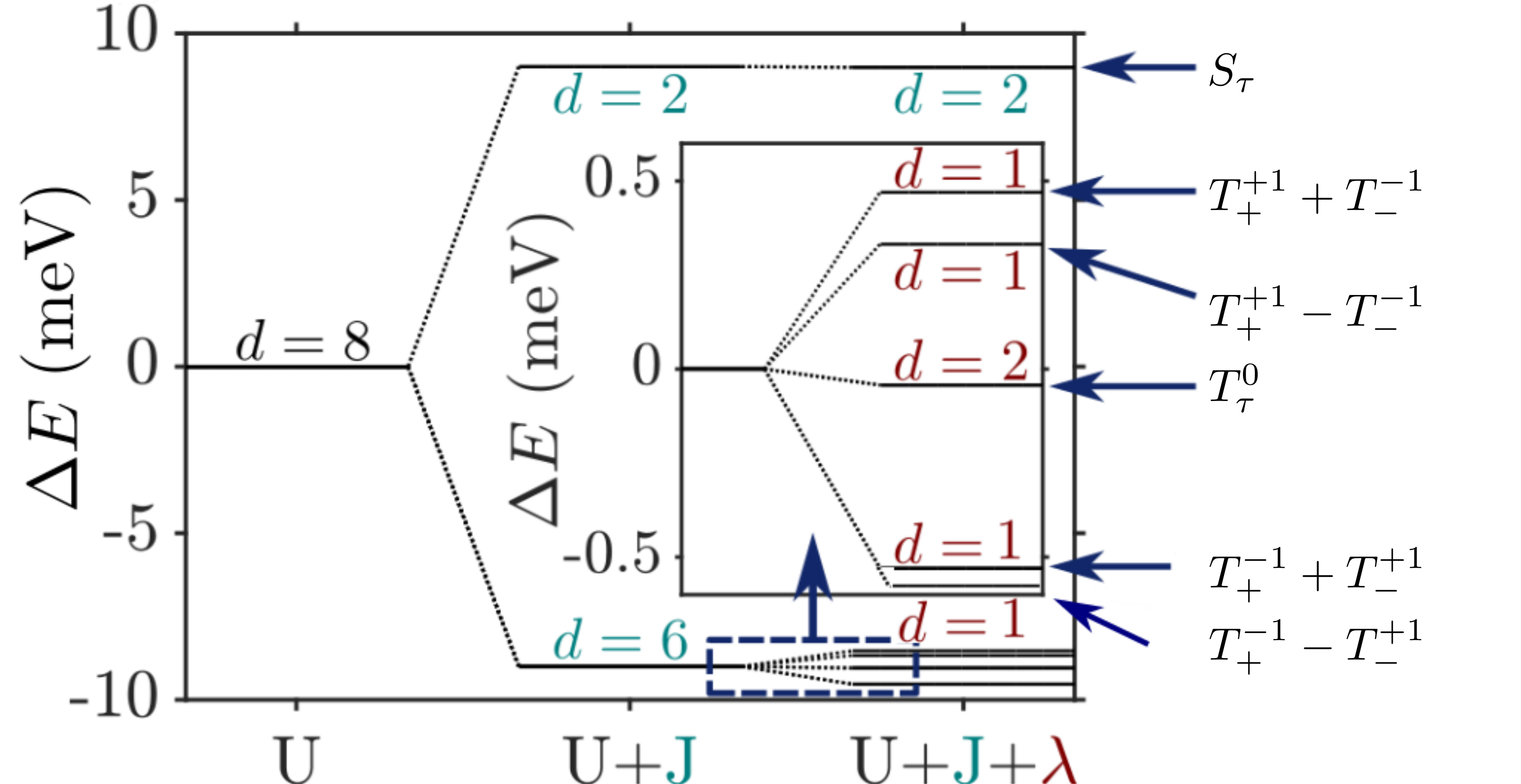}
    \caption{Low energy spectrum of CuPc. With only direct Coulomb interaction the spectrum is eightfold degenerate.
    Exchange coupling splits the triplet from the singlet states. The SOI further lifts the degeneracy of the triplet states.
    Adapted from Ref.~\cite{Siegert2015a}.}
    \label{fig:spectrum}
\end{figure}

The analysis of the anionic spectrum moves from the identification of three distinct energy scales in the system, i.e the direct Coulomb interaction ($U$, for simplicity), the exchange couplings ($J$) and the SOI parameters ($\lambda$), with the clear separation $U>J>\lambda$. The degeneracy, as well as the spin of the ground state is influenced by this three energy scales. 

Only considering the direct Coulomb interactions between the molecular orbitals would yield the degenerate anionic ground state
\begin{equation}
\ket{4\:\tau\sigma\sigma'} = \op[S\sigma]{d}^\dag\op[L\tau\sigma']{d}^\dag\ket{\Omega},
\end{equation}
with the eight-fold degeneracy ensured by the combination of the orbital degree of freedom $\tau$, the spin degree of freedom spin $\sigma$ in the SOMO and $\sigma'$ in the LUMO. The high Hubbard energy of the SOMO is responsible for the occupation of a LUMO, instead of the double occupation of the SOMO, in the anionic state.

By unitary transformation we can express these states also as two sets of singlets and triplets
\begin{equation}
\label{eq:basis}
    \begin{aligned}
    &\ket{S_\tau} = \frac{1}{\sqrt{2}}\left(\op[S\uparrow]{d}^\dag\op[L\tau\downarrow]{d}^\dag-
    \op[S\downarrow]{d}^\dag\op[L\tau\uparrow]{d}^\dag\right)\ket{\Omega},\\
    &\ket{T^{+1}_{\tau}} = \op[S\uparrow]{d}^\dag\op[L\tau\uparrow]{d}^\dag\ket{\Omega},\\
    &\ket{T^0_\tau} = \frac{1}{\sqrt{2}}\left(\op[S\uparrow]{d}^\dag\op[L\tau\downarrow]{d}^\dag+
    \op[S\downarrow]{d}^\dag\op[L\tau\uparrow]{d}^\dag\right)\ket{\Omega},\\
    &\ket{T^{-1}_\tau} = \op[S\downarrow]{d}^\dag\op[L\tau\downarrow]{d}^\dag\ket{\Omega},    
    \end{aligned}
\end{equation}
where the two sets, labeled by the angular momentum $\tau$, arise due to the degenerate LUMOs. We identify the orbital degree of freedom in the subscript, while the superscript of the triplet states gives the value of $S_z$. 
The exchange interaction  ($\propto J^{\mathrm{ex}}_\mathrm{SL}$) lifts the degeneracy between the triplet and singlet states, with the triplets becoming the new (still orbitally degenerate) ground state.
Eventually, the triplet-state degeneracy is partially lifted by the SOI which leads us to the actual eigenstates of the molecule, 
ordered from lowest to the highest in energy $\ket{T^{-1}_+} - \ket{T^{+1}_-}$, $\ket{T^{-1}_+} + \ket{T^{+1}_-}$, 
$\ket{T^{0}_\tau}$, $\ket{T^{+1}_+} - \ket{T^{-1}_-}$, and $\ket{T^{+1}_+} + \ket{T^{-1}_-}$. Time-reversal symmetry ensures the robustness of the singlets orbital degeneracy under the perturbation of the SOI: they only experience a slight downshift in energy of $\approx \SI{4e-2}{\meV}$.

The anionic low energy spectrum and its relation to the different energy scales associated with the molecule is depicted in \cref{fig:spectrum}. The splitting of $\approx \SI{1}{meV}$ of the triplet states due to the SOI is comparable, as we will discuss in the next section, to the energy broadening introduced by the coupling to the tip and substrate, thus making these states quasidegenerate, an essential requirement for the interference effects \cite{Schultz2009,Darau2009,Nilsson2010,Donarini2019} underlying the preparation and read-out protocols proposed here.

\subsection{Lightwave-STM}
\label{sec:lightwave-STM}

\begin{figure}
\begin{center}
    \includegraphics[width=0.8\linewidth]{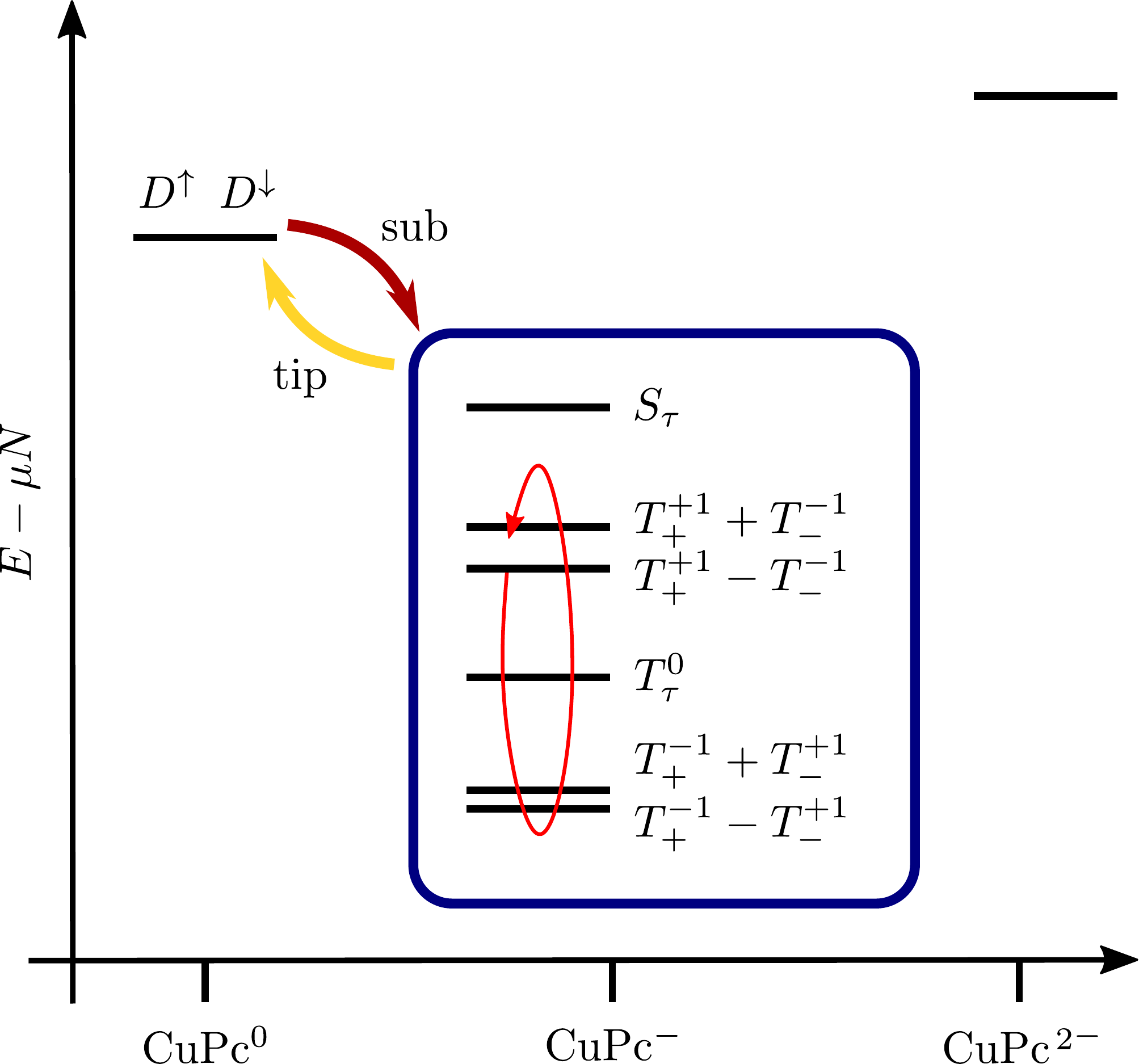}   
\end{center}
\caption{Schematic representation of the lightwave-STM dynamics within the many-body spectrum of CuPc. With the choice of chemical potential $\mu = \SI{-4}{eV}$, the low-energy anionic states have a lower energy than the ground states of the contiguous neutral and doubly charged configurations. The bias pulses induce tunneling transitions to the doublet states. The red arrow represent the SOI-induced precession. 
}
\label{fig:mb_spectrum}
\end{figure}

We describe the STM nanojunction in \cref{fig:setup} by means of the system-bath Hamiltonian 
\begin{equation}
\label{eq:hamil_general}
    \hamil = \hamil[M]+\hamil[IC]+\hamil[T]+\hamil[L],
\end{equation}
with the many-body molecular Hamiltonian $\hamil[M]$ introduced in the previous section.
The leads and the insulating thin film are responsible of polaronic and image charge effects\cite{Kaasbjerg2011,Olsson2007, Hernangomez2020} which renormalize the energies of the charged states. We simply account for this renormalization via the term 
\begin{equation}
    \hamil[IC] = -\delta_{\rm IC}(\op{N}-3)^2,
\end{equation}
with $\op{N}$ being the operator counting the total number of electrons occupying the frontier orbitals (three in the neutral state), and $\delta_{IC} = \SI{0.32}{\eV}$. This parametrization is obtained by fitting  our model to experimental transport gaps obtained from differential conductance measurements of CuPc on substrates with different work functions \cite{Siegert2016}. 
The molecule exchanges particles with the leads via a weak tunneling coupling. The state of the molecular junction in thermal equilibrium is thus obtained by considering a grand canonical ensemble with the equilibrium chemical potential of the leads. We choose such chemical potential to be $\mu = \SI{-4}{\eV}$, which keeps the molecule in its anionic state, i.e., with one additional electron, as shown schematically in \cref{fig:mb_spectrum}.

The leads are noninteracting Fermi seas with the Hamiltonian 
\begin{equation}
\hamil[L] = \sum_{\eta \bm{k}\sigma} [\epsilon^\eta_{\bm{k}} + \alpha_\eta V_{\rm bias}(t)] \op[\eta\bm{k}\sigma]{c}^\dag
    \op[\eta\bm{k}\sigma]{c},
\end{equation}
where $\eta$ labels the tip or substrate and $\op[\bm{k}\sigma]{c}$ destroys an electron with momentum $\bm{k}$ and spin $\sigma$ in the lead $\eta$. The spectrum of the lead is rigidly shifted by the applied bias across the junction. The same happens to the corresponding chemical potential $\mu_\eta(t) = \mu + \alpha_\eta V_{\mathrm{bias}}(t)$, thus ensuring the charge neutrality of the lead. Conventionally, we assume the energy levels of the molecule to remain unchanged during the pulse, while the tip and substrate levels are shifted in opposite directions \cite{Datta1997}. In our set up the bias drops asymmetrically across the junction, and the parameters $\alpha_\eta$, which depend on the tip height, are estimated as $\alpha_\mathrm{tip} = -0.54$ and $\alpha_\mathrm{sub} = 0.08$, according to electrostatic considerations \cite{Siegert2016}. The residual potential drop across the molecule is compensated by a polarization of the molecule \cite{Brandbyge2002}, whose influence on the molecular eigenstates we neglect in this work.

 It is the laser pulse impinging on the molecule to generate, in the lightwave-STM, a bias pulse across the junction \cite{Cocker2016, Peller2020}. Consequently, the tip and substrate chemical potentials acquire a time dependence which drives the system dynamics. The many-body eigenenergies together with the potential drop fractions $\alpha_\eta$ at the different leads set the thresholds biases at which new tunneling events can occur \cite{Donarini2010}, with the sensitivity set by the temperature. To a first approximation we give much more importance to the time interval of an open transition as to the exact shape of the bias pulse. We thus replace in our simulations the bias pulses by (smoothed) square pulses.      
  
The coupling between the molecule and the leads is given by the tunneling Hamiltonian
\begin{equation}
\hamil[T] = \sum_{\eta i\bm{k}\sigma} t^{\eta}_{i\bm{k}\sigma}\op[\eta\bm{k}\sigma]{c}^\dag
\op[i\sigma]{d}+ \mathrm{H.c.},
\end{equation}
with $\{t^\eta_{i\bm{k}\sigma }\}$ being the set of tunneling amplitudes. The latter are, according to the tunneling theory of Bardeen \cite{Bardeen1961}, the overlap integrals between the leads states (labeled here by momentum ${\bf k}$ and spin $\sigma$) and the molecular orbitals ($i = S,H,L\pm$ and the same spin $\sigma$). They encode the geometry of the contact, which, in a STM junction, is very asymmetric, with the atomically flat substrate opposed to the extremely sharp tip. This qualitative difference between the two contacts is clearly reflected into the single-particle tunneling rate matrices. The latter extend the concept of tunneling rates between energy levels in presence of degeneracies in the spectrum of the system \cite{Sobczyk2012,Donarini2019}. They are defined on the single-particle basis of the molecule as

\begin{equation}
    \Gamma^\eta_{i\sigma, j\sigma'}(E) = \frac{2\pi}{\hbar} \sum_{\bm{k}} (t_{i\bm{k}\sigma}^\eta)^*
    t_{j\bm{k}\sigma'}^\eta \delta(\epsilon_{\bm{k}}^\eta-E).
\end{equation}

In Ref.~\cite{Sobczyk2012} we calculate explicitly the tunneling rate matrices for a benzene molecule, starting from microscopic models for the leads. In this work we opt for a simplified form, still capturing, though, the fundamental symmetries of the problem. In particular, the tip's tunneling rate matrix must yield the topography of the molecular orbitals in the transport characteristics. To this end, in accordance with Chen's derivative rule \cite{Chen1990} for an $s$-symmetric tip, the tunneling matrix  assumes, for the generic element
\begin{equation}
    \label{eq:gamma_tip}
    \Gamma^{\mathrm{tip}}_{i\sigma,j\sigma'} (\bm{r}_\mathrm{tip}) = 
    \gamma^\mathrm{tip}_0 \psi^*_i(\bm{r}_\mathrm{tip})\psi_j(\bm{r}_\mathrm{tip}) \delta_{\sigma\sigma'},
\end{equation}
where $\gamma^\mathrm{tip}_0$ is proportional to the local density of states of the tip and we approximate it as a constant in the energy range relevant for our calculations (wide-band approximation).

On one end the wide-band approximation is a common limit taken in the mesoscopic transport community, unless the band structure of the leads is specifically on focus for the understanding of the experiment. On the other end, the local density of states (DOS) of a metallic tip is essentially an unknown function, due to its dependence on the microscopic configuration of the latter. It is common practice, in STM experiments, to select tips with a featureless DOS, in order to avoid spectroscopic signals, which do not stem from the electronic structure of the sample.

Moreover, we estimate the tip rate from a typical constant current set point of STM experiments on CuPc together with the corresponding topographical images \cite{Uhlmann2013}.
At a distance of $\SI{3}{\AA}$ from the horizontal symmetry plane of the molecule, we obtain a maximal tunneling rate to the LUMO orbitals $\max_{{\bm r}_{\rm tip}} \hbar\Gamma^{\rm tip}_{L\sigma L\sigma}({\bm r}_{\rm tip}) = \SI{0.4}{\meV}$. 
The localized tip breaks, in general, the rotational symmetry of the molecule and ensures the presence of off-diagonal terms in the associated tunneling matrix, when written in the energy and angular-momentum basis. The positivity of the rate matrix is instead guaranteed, in our approximation, by the fact that the tip tunneling matrix has by construction only a single nonzero eigenvalue, while its trace is proportional to the sum over the orbital densities calculated at the tip position. The frontier orbitals S, H and L$\pm$ belong to different irreducible representations of the $C_{4v}$ symmetry group characterizing CuPc on the NaCl substrate. The substrate tunneling rate matrix is thus diagonal. The microscopic proof of this statement is found in Ref.~\cite{Sobczyk2012} where the analogous case of a benzene molecule has been analyzed in detail. For simplicity we reduce here the substrate tunneling rate matrix to the form  
\begin{equation}
    \label{eq:gamma_sub}
    \Gamma^{\mathrm{sub}}_{i\sigma,j\sigma'}  = \Gamma^\mathrm{sub}_0  \delta_{ij} \delta_{\sigma\sigma'},
\end{equation}
thus neglecting the differences between the molecular orbitals. The diagonal components associated with the two LUMOs are identical due to time-reversal symmetry. Moreover, the tunneling events connecting the neutral and the anionic low-energy states, considered here, mostly involve the addition or removal of an electron in the LUMOs and  justify, for the sake of simplicity, the approximation in Eq.~\eqref{eq:gamma_sub}. The overall strength of the tunneling coupling to the metallic substrate has been fixed to $\hbar \Gamma^\mathrm{sub}_0 = \SI{1}{\meV}$, roughly in accordance with DFT calculations \cite{Repp2005} with a bilayer of NaCl on Cu(111).

\subsection{Transport theory}
\label{sec:transport}
The electronic transport calculation reported in this work is based on the STM theory for single molecules on thin insulating films presented in Ref.~\cite{Sobczyk2012}. We proceed in the framework of a generalized master equation, which naturally allows for the treatment of strong electronic correlations on the molecule \cite{Timm2008, Donarini2012, Gaudenzi2017, Yu2017}. The latter plays a crucial role in STM on thin insulating film, because the presence of the insulator hinders the screening associated with the hybridization with the metallic substrate and enhances the specific features of the pristine molecule \cite{Repp2005, Repp2010, Schulz2015, Yu2017}.
The starting point is the Liouville-von Neumann (LvN) equation for the full density matrix 
\begin{equation}
\dot{\rho}_\mathrm{tot} = -\tfrac{i}{\hbar} \comm{\hamil(t)}{\op[\rm tot]{\rho}},
\end{equation}
where, due to the laser-induced bias pulses, the explicit time dependence in the Hamiltonian has been highlighted.  
Because we treat the tunneling perturbatively, it is convenient to write the LvN equation in the interaction picture $ \dot{\rho}^I_\mathrm{tot} = -\frac{i}{\hbar} \comm{\hamil[T]^I}{\op[\rm tot]{\rho}^I}$. Moreover, we are primarily interested in the dynamics of the molecule, thus we aim at the equation of motion for the reduced density matrix (RDM), obtained by tracing out the lead degrees of freedom $\rho = \Tr[L]{\op[\rm tot]{\rho}}$. The initial time factorization for the density operator into a molecule and a lead component $\op[\rm tot]{\rho}(t_0) = \op[\rm M]{\rho} (t_0)\otimes \rho_{\rm L}(t_0)$ allows one to derive the following integro-differential equation of motion for the reduced density matrix
\begin{equation}
\label{eq:GME_I}
    \dot{\rho}^I(t) = \int_{t_0}^t \mathcal{K}^{(2)}_I(t,t')\rho^I(t'){\rm d}t',
\end{equation}
valid to second order in the tunneling coupling, with the interaction picture propagation kernel 
\begin{equation}
\begin{split}
  \mathcal{K}&^{(2)}_I(t,t')\rho^I(t') = \\ 
&-\frac{1}{\hbar^2}
  \Tr[L]{\comm{\hamil[T]^I(t)}{\comm{\hamil[T]^I(t')}{\rho^I(t') \otimes \op[\rm L]{\rho}(t_0)}}}.
\end{split}
\end{equation}

In Eq.~\eqref{eq:GME_I} the dynamics of the RDM at time $t$ can be  determined only with the knowledge of the RDM at all previous times, starting at the initial time $t_0$, thus showing a memory effect. The propagator kernel vanishes, though, for $t-t' \to \infty$, due to its proportionality to the lead correlator
\begin{equation}
    F(t,t';\eta) = \sum_{\bm{k}\sigma} \Tr[L]{\op[\eta\bm{k}\sigma,I]{c}^\dagger(t)\op[\eta\bm{k}\sigma,I]{c}(t') \rho_{\rm L}(t_0)}.
\end{equation}
The time-scale $\tau_{\rm rel}$ of its exponential decay, is given, for a thermal bath of noninteracting fermions, by $\tau_{\rm rel} \approx \hbar \beta$ \cite{Donarini2012b} where $\beta = (k_BT)^{-1}$ is the inverse thermal energy. The speed in the dynamics of the RDM is instead measured by $\hbar\Gamma$, the overall level broadening introduced by tunneling and by a third time-scale associated with the variation speed of the bias pulse $\tau_{\rm pulse}$. We concentrate on the weak-coupling, adiabatic limit $1/\Gamma \approx \tau_{\rm pulse} \gg \tau_{\rm rel}$. 

Within this limit, both the RDM $\rho^I$ and the electrochemical potential can be evaluated in Eq.~\eqref{eq:GME_I} at the time $t$. Such an approximation is justified since the lead correlation function decays much faster than the time-scale for the bias modulation and the RDM dynamics. In our choice of parameters, the relaxation time $\tau_{\rm rel}$ is much smaller than the tunneling time $1/\Gamma$ and the pulse duration. Similarly to Ref.~\cite{Donarini2012b}, we obtain, from Eq.~\eqref{eq:GME_I}, the time local propagator kernel
\begin{equation}
\begin{aligned}
\label{eq:GME_Markov}
    \dot{\rho}^I(t) = &\frac{-1}{\hbar^2} \int_{0}^{\infty}dt'\: \\
    &\Tr[L]{
\comm{\hamil[T]^I(t)}{\comm{\hamil[T]^I(t-t')}{\rho^I(t)\otimes \rho_{\rm L}(t_0)}}}.
\end{aligned}
\end{equation}
Moreover, vibrations in the contacts reduce the relaxation time of the electronic correlator, by introducing additional dissipative channels for the electronic excitations, as was recently discussed in Refs.~\cite{Thomas2019,Peskin2019}. Thus the range of validity of our approximation should also accommodate the cryogenic temperatures and tunneling couplings of the lightwave-STM experiments \cite{Cocker2016, Peller2020}. 

In support of our argument we notice, moreover, that the differential conductance peaks in STM experiments on thin insulating films \cite{Repp2005, Schulz2015, Repp2010, Pavlicek2013} show a Gaussian profile and a width clearly much larger than both the thermal energy $k_BT$ and the tunneling-induced broadening $\hbar\Gamma$. The role of substrate optical phonons in the explanation of such anomalous spectral broadening of the conductance has been demonstrated \cite{Repp2005b} and also controlled by changing the insulating layer from a NaCl to a RbI or Xe one \cite{Pavlicek2013}. 

By evaluating the time integrals in Eq.~\eqref{eq:GME_Markov} and converting the equation to the Schr\"odinger picture, we finally obtain
the desired generalized master equation (GME):
\begin{equation}
    \label{eq:GME}
    \dot{\rho}(t) =   [\liou[M] +  \liou[T](t)+ \liou[LS](t)]\rho(t).
\end{equation}

The first term $\liou[M]$ describes the coherent evolution of the isolated molecule. For later convenience we express it in the Liouville superoperator formalism
\begin{equation}
\liou[M] = - \frac{\iu}{\hbar} \sum_{\alpha=\pm 1} \alpha \mathcal{H}_{\rm M,\alpha},
\end{equation}
where $\mathcal{H}_{\rm M,\alpha}$ acts on the following density operator as,
$\mathcal{H}_{\rm M,+}\rho := \hamil[M]\rho$ and $\mathcal{H}_{\rm M,-}\rho = \rho\hamil[M]$.

The tunneling coupling between the system and the leads treated in the lowest nonvanishing perturbative order is responsible for the two additional terms in Eq.\eqref{eq:GME}. We refer to them as the tunneling and the Lamb-shift Liouvillean, respectively. In the superoperator notation, they read
\begin{widetext}
\begin{equation}
\label{eq:L_T}
    \liou[T](t) = - \frac{1}{2} \sum_{\alpha_1\alpha_2}\sum_{nm}\sum_{p\eta} 
    \alpha_1\alpha_2\Gamma^{\eta,p}_{n,m}  
    \mathcal{D}_{n,\alpha_2}^{\bar{p}} 
    f_\eta^{(\alpha_1 p)}(\iu\hbar p\liou[0])\mathcal{D}_{m,\alpha_1}^{p},
\end{equation}
and
\begin{equation}
\label{eq:L_LS}
    \liou[LS](t)= -\frac{\iu}{2\pi} \sum_{\alpha}\sum_{nm}\sum_{p\eta} 
    \alpha \Gamma^{\eta,p}_{n,m} 
    \mathcal{D}_{n,\alpha}^{\bar{p}} p_\eta(\iu\hbar p\liou[0]) \mathcal{D}_{m,\alpha}^{p}.
\end{equation}
\end{widetext}

The tunneling and the Lamb shift Liouvilleans both depend on time via the lead chemical potential $\mu_\eta(t)$ which shifts the argument of the Fermi function $f_\eta^q(\epsilon) := \left\{1+\e^{q\beta[\epsilon-\mu_{\eta}(t)]}\right\}^{-1}$ and of the principal part function $p_\eta(\epsilon) = -\Re\psi\{\tfrac{1}{2} + \tfrac{\iu\beta}{2\pi}[\epsilon-\mu_\eta(t)]\}$, the latter being defined via the real part of the digamma function $\psi$. The collective indices $n$ and $m$ fully identify the molecular orbital, while $p = \pm$ and $\bar{p} = -p$ distinguish in $\mathcal{D}^p_{n,\alpha}$ the creation from the annihilation operator, associated respectively to $p = +1$ and $p = -1$. Notice, moreover, that $\Gamma^{\eta,-}_{n,m} = \Gamma^{\eta}_{n,m}$ and $\Gamma^{\eta,+}_{n,m} = \Gamma^{\eta}_{m,n}$. Analogously to the coherent Liouvillean presented above, the creation and annihilation operators are dressed with the Liouville index $\alpha$, which identifies from which side they act on the following operator. 

Additionally, we only include in Eq.~\eqref{eq:L_T} and \eqref{eq:L_LS} the unperturbed Liouvillean of the system, $\liou[0] = - \frac{\iu}{\hbar} \sum_{\alpha'=\pm 1}\alpha' \mathcal{H}_{0,\alpha'}$, in the calculation of $\mathcal{L}_{\rm T}$ and $\mathcal{L}_{\rm LS}$. The energy splitting $\delta\epsilon$ induced by the SOI is comparable to the level broadening $\hbar\Gamma$.  We perform, however, a perturbation expansion up to first order in $\Gamma$ and therefore we must exclude from the dynamics terms proportional to $\hbar \Gamma \delta\epsilon$ which are comparable to cotunneling contributions $O(\Gamma^2)$.  
This approach is equivalent to treating the problem in the singular-coupling limit, as discussed in Refs.~\cite{Spohn1980, Schultz2009}.

The vacuum barrier and the thin insulating film separating the molecule respectively from the tip and the metallic substrate ensure a weak tunneling coupling to the electrodes. The perturbative expansion in the tunneling is justified by the combination of such a weak tunneling coupling  with a large Coulomb charging energy, and the temperature higher than the Kondo temperature ($T_K$). The lowest perturbative order considered in this paper well describes single-electron sequential tunneling events between the tip and the molecule or the molecule and the substrate, as triggered by over-threshold bias pulses. Higher-order expansion in the tunneling coupling, would also capture cotunneling events connecting directly the tip and the substrate with only virtual charge fluctuations on the molecule. In particular, inelastic cotunneling could be relevant to characterize under-threshold bias pulses. Conversely, in the Kondo regime ($T < T_K$), the perturbative approach breaks down completely.

While the tunneling events connecting many-body eigenstates with adjacent particle number are described by $\mathcal{L}_{\rm T}$, the  Lambd-shift Liouvillean $\mathcal{L}_{\rm LS}$ outlines the virtual transitions of electrons which preserve the particle number on the molecule, with the net result of renormalizing the coherent dynamics generated by the system Hamiltonian. In this sense it is possible to write the Lamb shift Liouvillean as a commutator with an effective system Hamiltonian $\hamil[LS]$. 

For a lightwave-STM in the weak tunneling coupling regime, the fundamental role of the bias is to trigger sequential tunneling events, i.e., for the system under study, the over-threshold bias opens the transition between the anionic and the neutral states of the molecule. The bias pulse acts in our simulations as a time-dependent shift of the chemical potentials in the leads. Furthermore, such a shift enters the tunneling Liouvillean as an argument of Fermi functions, which rapidly switch on and off when the bias crosses a threshold value. Therefore, independently of its actual shape, every threshold-trespassing  pulse acts, at low temperatures, as a square pulse.\\
Due to the logarithmic tails of the digamma functions, the effect of the bias shape is more subtle in the Lamb shift contribution to the dynamics. The latter, as we will see, mainly influences, though, the free evolution of the molecule (with constant zero bias) and plays only a marginal role during the dynamically driven preparation and read-out.

The SOI-induced evolution renormalized by the Lamb shift and the tunneling are characterized by comparable time-scales. They combine to yield a driven nonadiabatic evolution of the system. The reduced density matrix fully captures such dynamics. We propose, as a read-out mechanism, the measurement of the average current flowing through the STM nanojunction due to repeated pump-probe cycles, recorded as a function of the delay time between the pulses. To this end we first calculate the time-dependent reduced density matrix by solving the GME in Eq.~\eqref{eq:GME} with a thermal initial condition. The latter is a reasonable assumption for sufficiently long repetition periods of the pump-probe cycle. The current through the system is than obtained as

\begin{equation}
    \label{eq:current}
    \langle I_\eta\rangle (t) = \Tr[M]{[\mathcal{J}^+_\eta(t) - \mathcal{J}^-_\eta(t)] \rho(t)},
\end{equation}
where the time-dependent jump superoperators $\mathcal{J}^\pm_\eta(t)$ refer to the tunneling events to and from the lead $\eta$ respectively. They read
\begin{equation}
\begin{split}
    \mathcal{J}^+_\eta(t) &= \sum_{n,m}\Gamma_{n,m}^\eta \mathcal{D}^-_{m,+}f_\eta^-(\iu\hbar\liou[0])\mathcal{D}^+_{n,-},\\
    \mathcal{J}^-_\eta(t) &= \sum_{n,m}\Gamma_{n,m}^\eta \mathcal{D}^-_{m,-}f_\eta^+(\iu\hbar\liou[0])\mathcal{D}^+_{n,+},
\end{split}
\end{equation}
where the explicit time dependence is contained, as for the tunneling and Lamb shift Liouvilleans, in the time-dependent chemical potentials which shift the argument of the Fermi functions.

\subsection{Operator space}
\label{sec:Operator_space}
The density matrix $\rho$ describing an $n$-dimensional Hilbert space is, in general, a $n \times n$ Hermitian, positive definite matrix, thus parametrized by $n^2$ real numbers.\footnote{One of these numbers is always fixed to $1 = \Tr{\rho}$, due to the probabilistic nature of the density matrix.} This number of parameters is the size of the Liouville (vector) space to which $\rho$ belongs. By exploiting the notion of scalar product, also defined on Liouville spaces, we obtain, for any orthogonal basis set $\{B_i\}$, the expansion    
\begin{equation}
\label{eq:Liou_exp}
    \rho = \sum_i \frac{\langle\langle B_i| \rho \rangle\rangle B_i}{\langle\langle B_i|B_i\rangle\rangle},
\end{equation}
with $|\rho \rangle\rangle$ the vector notation for the density operator in Liouville space and the scalar product defined as $\langle\langle A|B\rangle\rangle=\Tr{A^\dag B}$. 
Interestingly, the scalar product $\langle\langle B_i| \rho \rangle\rangle \equiv \langle B_i \rangle$ is equivalent, for a basis of  Hermitian operators, to the expectation value of the observable $B_i$. Thus, the combination of Eqs. \eqref{eq:GME_Markov} for a suitable basis set and \eqref{eq:Liou_exp} translates the GME into a set of coupled linear differential equations for $n^2$ expectation values of system observables. Those equations give valuable information about the interplay of molecular observables and a more intuitive physical picture of the molecular dynamics with respect to the original GME.

As already discussed, we are interested into the coupled dynamics of the neutral ground state and of the low-enerrgy set of triplet and singlet anionic states. The corresponding density matrix consists of three separate blocks, because (i) coherences between states with different particle number are forbidden in absence of superconducting correlations and (ii) the energy splitting between the singlet and the triplet exceeds by far the tunneling coupling, thus justifying the use of the secular approximation which neglects the fast oscillating coherences between singlet and triplet states.

The matrix block associated with the neutral, spin degenerate state can be expanded on the basis 
$\mathcal B^{\nicefrac{1}{2}} = \{\mathbb{1}_2, \tfrac{\op[x]{\sigma}}{2}, \tfrac{\op[y]{\sigma}}{2},\tfrac{\op[z]{\sigma}}{2}\}$, with $\sigma_\alpha$ the Pauli matrices, to obtain
\begin{equation}
    \rho^{\rm d} = \frac{p^{\rm d}}{2}\mathbb{1}_2 + \langle{\bf S}^{\rm d}\rangle\cdot\boldsymbol{\sigma},
\end{equation}
where $p^{\rm d}$ is the population of the doublet neutral level and $S^{\rm d}_\alpha = \sigma_\alpha/2$ is the matrix representation of the $\alpha$ component of the spin operator in the doublet basis and vanishes elsewhere. 

The orbitally degenerate singlet anionic states are treated analogously. The definition of a set of pseudospin operators $\tau^{\rm s}_\alpha = \sigma_\alpha/2$ with $\alpha = x,y,z$ leads to the analogous decomposition 
of the corresponding block in the density matrix
\begin{equation}
    \rho^{\rm s} = \frac{p^{\rm s}}{2}\mathbb{1}_2 + \langle \bm{\tau}^{\rm s}\rangle\cdot\boldsymbol{\sigma},
\end{equation}
where $p^{\rm s}$ is the total population of the singlet degenerate level. 
The most interesting sub-block is the one spanned by the triplet states. The triplet space is of the tensor product of two spaces, respectively a spin 1 and a pseudospin $\nicefrac{1}{2}$ which accounts for the additional orbital degeneracy. 
The spin operators associated with a generic spin 1 system are represented in the basis of the $S_z$ eigenvectors by the matrices 
\begin{equation}
\begin{split}
    &S_x = 
    \frac{1}{\sqrt{2}}
    \begin{pmatrix}
        0&1&0\\
        1&0&1\\
        0&1&0
    \end{pmatrix},
    \,
    S_y =
    \frac{1}{\sqrt{2}}
    \begin{pmatrix}
        0&-\iu&0\\
        \iu&0&-\iu\\
        0&\iu&0
    \end{pmatrix}\\
&\phantom{m}\\
&\text{and} \quad
S_z = 
\begin{pmatrix}
 1 & 0 & 0\\
 0 & 0 & 0\\
 0 & 0 & -1
\end{pmatrix}.
\end{split}
\end{equation}

Given that the Hilbert space is three dimensional, the associated Liouville space has dimension nine.
Therefore, it is not enough to expand the corresponding density matrix in terms of the identity and the spin operators. Five additional operators are needed to complete a basis set. The spin quadrupole operators do the job:
\begin{equation}
    \begin{aligned}
   &S_{z^2} = 2S_z^2-S^2_x-S^2_y ,\\
   &S_{xz} = S_xS_z + S_zS_x,\\
   &S_{yz} = S_yS_z+S_zS_y,\\
   &S_{x^2y^2} = S_x^2-S_y^2,\\
   &S_{xy} = S_xS_y+S_yS_x.
    \end{aligned}
\end{equation}
Altogether we set up the following basis for the description of a spin 1 density matrix 
\begin{equation}
    \mathcal{B}^1 = \bigg\{\mathbb{1}_3, \op[x]{S},\op[y]{S},\op[z]{S},\op[z^2]{S},\op[xz]{S},\op[yz]{S},
    \op[x^2y^2]{S},\op[xy]{S}\bigg\}.
\end{equation}
By combining the two bases $ \mathcal{B}^{\nicefrac{1}{2}}$ and $ \mathcal{B}^1$ we obtain a basis in which we can expand the density matrix for the triplet space
\begin{equation}
    \label{eq:basis_trip}
    \mathcal{B}^\mathrm{t}= \mathcal{B}^1\otimes \mathcal{B}^{\nicefrac{1}{2}}.
\end{equation}
The description of the orbitally degenerate triplet space requires more than just the population $p^t = \Tr[]{\rho^t}$ and the expectation value of the separate spin (dipole and quadrupole) and pseudospin operators, calculated respectively as

\begin{equation}
    \begin{split}
        \langle S_i^{\rm t}\rangle = \Tr[]{\rho^{\rm t} \left(S_i\otimes \mathbb{1}_2\right)},
    \end{split}
\end{equation}
and 
\begin{equation}
           \langle \tau_\alpha^{\rm t}\rangle = \Tr[]{\rho^{\rm t} \left(\mathbb{1}_3\otimes \frac{\sigma_\alpha}{2}\right)},
\end{equation}
where $\rho^{\rm t}$ is the triplet block of the density matrix, $S_i$ is one of the spin operators in $\mathcal{B}^1$, and $\sigma_\alpha$ is one of the Pauli matrices. In general, also the mixed spin and pseudospin correlators 
\begin{equation}
    \begin{split}
        \langle S_i^{\rm t}\tau^{\rm t}_\alpha\rangle = \Tr[]{\rho^{\rm t} \left(S_i\otimes \frac{\sigma_\alpha}{2}\right)},
    \end{split}
\end{equation}
must also be taken into account. These mixed correlators play a crucial role in the description of SOI-induced dynamics, in which spin and orbital evolution are intertwined and the equation of motion of the spin or pseudospin variables is coupled to that of the correlators, since the latter do not factorize $p^{\rm t}\langle S_i^{\rm t}\tau^{\rm t}_\alpha\rangle \neq \langle S_i^{\rm t}\rangle \langle\tau^{\rm t}_\alpha\rangle$, with $p^{\rm t}$ the occupation of the triplet states. 

\section{Initial state preparation}
\label{sec:prep}
The pump-laser pulse impinging on the molecule excites it out of  its thermal equilibrium and is responsible for its initial-state preparation. The evolution of the molecular many-body state during the preparing pulse is characterized by the interplay of several processes giving rise to a complex dynamics. To scrutinize the role of the different contributions, we examine the initial-state preparation step by step. First we concentrate only on the interference effects associated with the tunneling processes. In a second and third steps we add the contributions of the SOI and the Lamb shift.
\subsection{Tunneling dynamics}
\begin{figure}
    \centering
    \includegraphics[width=\linewidth]{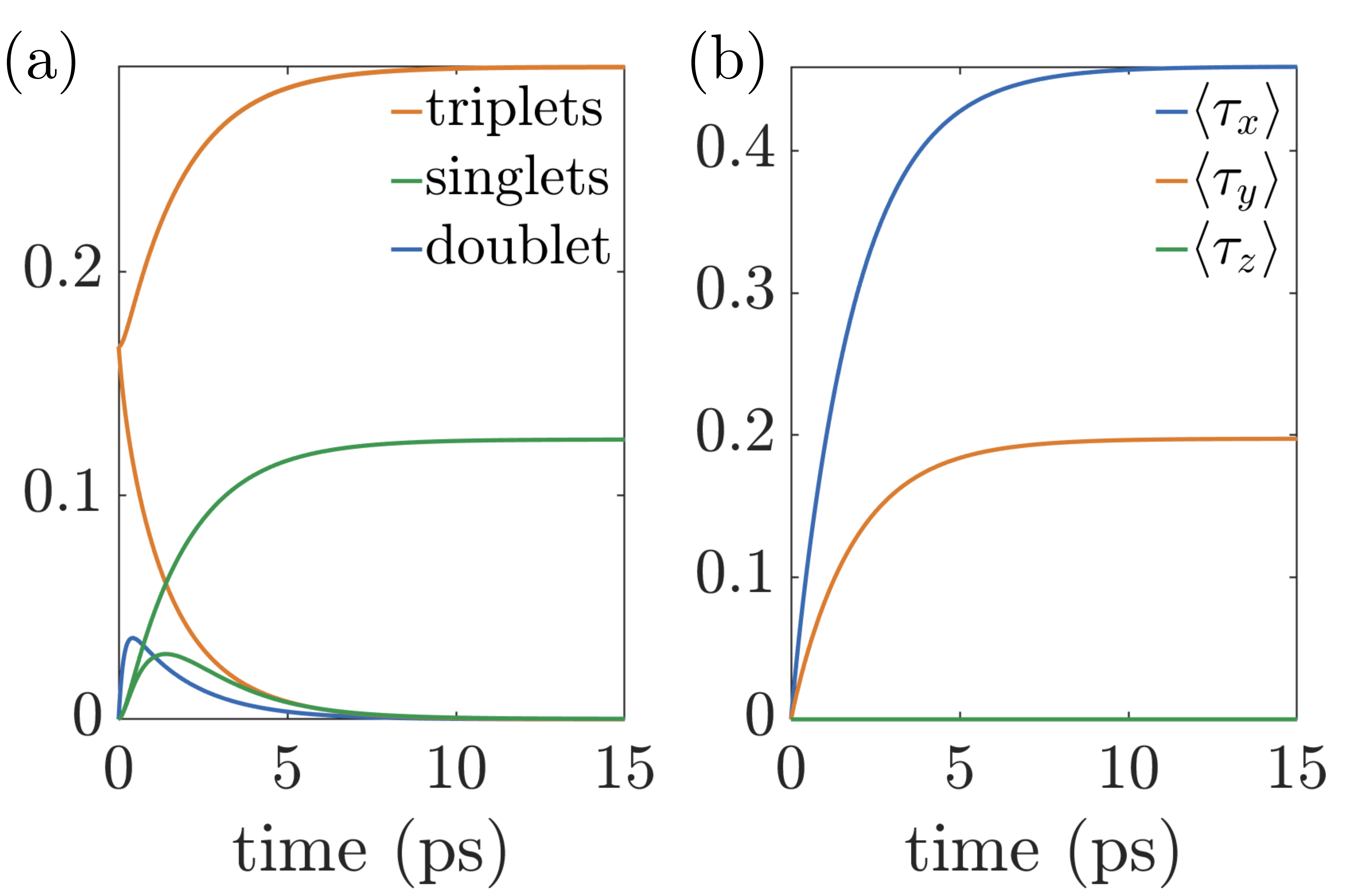}
    \caption{Blocking state in the high-bias limit. (a) Eigenvalues of the RDM plotted as a function of time. After a short excursion to  the neutral state, the system gets blocked in the anionic state. A threefold (twofold) degeneracy for the eigenvalues associated with the triplets (doublets) follows from spin isotropy. (b) Time evolution of the pseudospin components under the same conditions as in panel (a). The blocking state has nonzero expectation values for the pseudospin.}
    \label{fig:ev_tunneling}
\end{figure}

The fundamental mechanism underlying the initial-state preparation of the CuPc is the interference blocking characterizing nanojunctions with a quasidegenerate many-body spectrum and weakly coupled to the leads \cite{Michaelis2006,Donarini2010,Nilsson2010,Donarini2012}. This blockade, previously investigated only in the stationary limit, only arises when the applied bias voltage exceeds the threshold for triggering  sequential tunneling events and, moreover, unidirectional transport is induced through the junction (high-bias limit). 

\begin{figure}[h!]
    \centering
    \includegraphics[width=\linewidth]{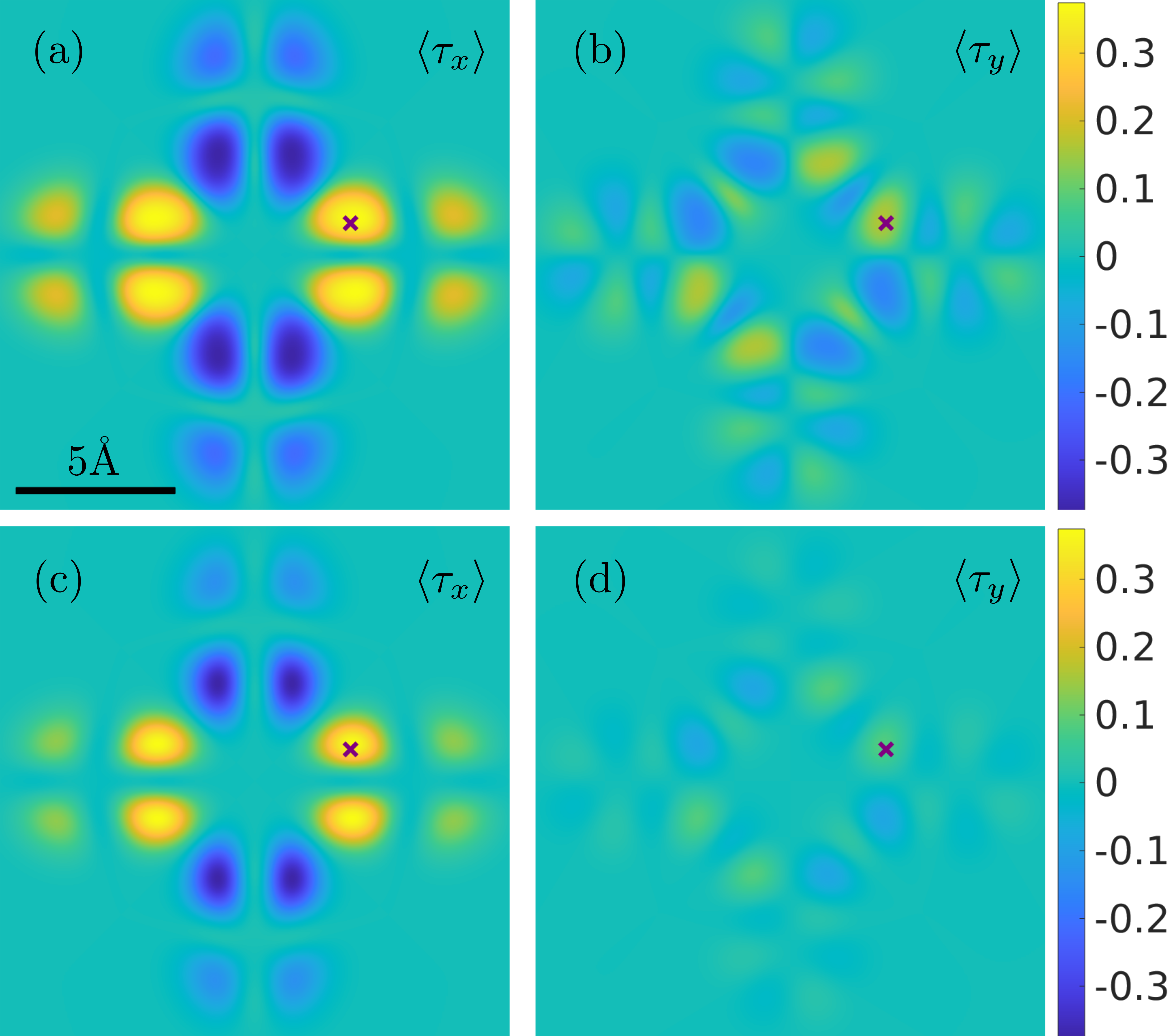} 
    \caption{Strong dependence of the dark state preparation on the tip position. We plot the $x$ and $y$ component of the pseudospin as a function of the tip position. In panels (a) and (b) we neglect SOI and the Lamb shift. 
    The SOI is introduced in panels (c) and (d). The black cross marks the tip position associated with the simulations presented in the rest of the paper. The SOI reduces the strength of the pseudospin obtained by the preparing pulse.}
    \label{fig:prep_tunneling}
\end{figure}

For this reason, we first consider the STM junction under a constant over-threshold bias and analyze how the system achieves the stationary state, as shown in \cref{fig:ev_tunneling}.  The tip position is fixed at $\SI{3}{\AA}$ above the horizontal symmetry plane of the molecule, in the vicinity of one of the lobes of the $L_{yz}$ orbital (compare \cref{fig:prep_tunneling} with \cref{fig:real_lumos}). Moreover, we apply a (sample) bias of $V_{\rm b}=-\SI{0.4}{V}$ to the system and, at first, neglect the SOI and Lamb-shift contribution to the dynamics. The RDM is thus governed by the differential equation $\dot{\rho} = \liou[T]\rho$ supplemented by the thermal initial condition $\rho(t_0) =  \exp[{-\beta(\hamil[M]-\mu N)}]/Z_{\rm G}$ with $Z_{\rm G}$ being the grand canonical partition function. 
The simulation (as also the others reported in this paper) has been performed at a temperature of $T = \SI{30}{\K}$. Initially, the system has an essentially equal probability to be in any of the triplet anionic states, while the singlets and the neutral doublet are empty. Thermal fluctuations are, in fact, not strong enough to populate these states, as $E^{\rm s} - E^{\rm t} \gg k_B T$ and also $E^{\rm d} - E^{\rm t} + \mu\gg k_B T$, being, $E^{\rm t}$, $E^{\rm s}$, and $E^{\rm d}$, respectively, the energy of the triplets, singlets, and doublet levels, calculated for simplicity in absence of the SOI perturbation.

The bias, applied here from the beginning, is instead large enough to trigger transitions between the anionic triplets and the neutral doublet ground state with one electron tunneling from CuPc towards the tip. Therefore, as can be seen from the eigenvalues of the density matrix plotted in \cref{fig:ev_tunneling} (a), at first the doublet gets populated. Soon, though, tunneling of an electron from the substrate brings the molecule back to an anionic configuration in which, on average, both singlets and triplets are populated. 

For a time lapse of a few picoseconds both the neutral and the anionic levels are populated and current flows through the molecule. Gradually some triplet states (three of them, in fact) and one singlet state act as probability sinks. Despite transitions to the neutral doublet remaining energetically available (the bias is kept constant, here), within approximately $\SI{4}{\ps}$ the entire occupation probability is concentrated in the anionic state and the current is blocked. This evolution of the density-matrix eigenvalues is the time-resolved observation of the dark-state formation as theoretically predicted \cite{Michaelis2006, Nilsson2010, Sobczyk2012, Darau2009} and recently observed in carbon nanotubes \cite{Donarini2019}. 

The more direct fingerprint of the interference-blocking mechanism is represented, though, by the nonzero expectation value of the pseudospin $x$ and $y$ components reached contemporary to the current blocking, as seen \cref{fig:ev_tunneling} (b). To understand the origin of this phenomenon, we consider the equation of motion for the triplet and singlet pseudospin vectors derived from $\dot{\rho} = \liou[T]\rho$. We obtain
\begin{equation}
\label{eq:pseudospin_tun}
    \begin{aligned}
        \frac{\rm d}{{\rm d}t}\ev{\bm{\tau}^\mathrm{s}} =&-\sum_\eta \gez f^-_\eta(\epsilon_\mathrm{s})\ev{\bm{\tau}^{\mathrm{s}}}\\
        &+ \frac{\gtz}{2}\left(\fptS p^\mathrm{d}-\fmtS p^{\mathrm{s}}\right) \bm{P}_\tau,\\
    \frac{\rm d}{{\rm d}t} \ev{\bm{\tau}^\mathrm{t}}=&-\sum_{\eta} \gez f^-_\eta(\epsilon_\mathrm{t}) \ev{\bm{\tau}^\mathrm{t}}\\
    &+ \frac{\gtz}{2}\left(3 \fpt p^\mathrm{d}-\fmt p^{\rm t}\right) \bm{P}_\tau,
\end{aligned}
\end{equation}
where $\gtz = \tfrac{\gamma_0^{\rm tip}}{2}(\psi_{xz}^2 + \psi_{yz}^2)$, $\gsz = \Gamma^{\rm sub}_0$, are the average rates to the tip and substrate, respectively. Moreover we introduced the energies $\epsilon_{\rm t/s} = E_{\rm t/s} - E_{\rm d} < 0$ and the vector $\bm{P}_\tau$
\begin{equation}
\label{eq:tip_pol}
    \bm{P}_{\tau} = \frac{1}{\psi_{xz}^2+\psi_{yz}^2} 
    \begin{pmatrix} \psi_{xz}^2-\psi_{yz}^2  \\ 
    2\psi_{xz}\psi_{yz}  \\ 0\end{pmatrix},
\end{equation}
whith $\psi_{xz}$ and $\psi_{yz}$  being the wave functions of the real valued LUMOs depicted in \cref{fig:real_lumos}.
The over-threshold bias with enforced unidirectional transport implies that $\fpt$,  $\fptS$, $\fms$ and $\fmsS$ are exponentially suppressed. Looking for the stationary solution of Eqs. \eqref{eq:pseudospin_tun}, under those assumptions, we obtain the following conditions 
\begin{equation}
\label{eq:pseudospin_stat}
        \ev{\bm{\tau}^\mathrm{t}} = -\frac{1}{2} \bm{P}_\tau p^\mathrm{t}, \quad  
        \ev{\bm{\tau}^\mathrm{s}} = -\frac{1}{2} \bm{P}_\tau p^\mathrm{s}, \\ 
\end{equation}
which, when inserted into the equations for the populations, yield
\begin{equation}
    \begin{aligned}
        \dot{p}^\mathrm{t} & = \sum_\eta [ 3\gez\fpe ]p^\mathrm{d} -\gsz\fms p^\mathrm{t},\\
        \dot{p}^\mathrm{s} & = \sum_\eta [ \gez\fpeS ]p^\mathrm{d} -\gsz\fmsS p^\mathrm{s},\\
        \dot{p}^\mathrm{d} &= \sum_\eta -\gez[3\fpe +\fpeS] p^\mathrm{d} \\&+
        \gsz\fmsS p^\mathrm{s}+\gtz\fms p^\mathrm{t}.
    \end{aligned}
\end{equation}
The equations above do not contain any rate proportional to $f_{\rm tip}^-(\epsilon_{\rm t/s})$, thus revealing how tunneling events towards the tip are completely suppressed.  

Thus, on a time-scale fixed by the smallest of the tip and the substrate rates, the neutral state gets depopulated, $p^{\rm t} + p^{\rm s} = 1$, and the total pseudospin reads
\begin{equation}
\label{eq:blocking_pseudo}
    \ev{\bm{\tau}} = -\frac{1}{2}\bm{P}_\tau.
\end{equation}

In close analogy to the spin-valve problem \cite{Braun2005} we interpret this condition as pseudospin accumulation in a direction antiparallel to the lead polarization $\bm{P}_\tau$, with the formation of an electronic dark state. The molecule is, in this configuration, completely decoupled from the tip and thus, in the infinite time limit, the ratio between the populations of the singlet and the triplet is expected to reduce to the Boltzmann factor. This condition, though, is clearly not the one shown in \cref{fig:ev_tunneling}, even after $\SI{15}{ps}$. The steady state would be reached only over a much longer time-scale, set by the depopulation rate of the singlet towards the substrate, which is strongly suppressed due to the Coulomb blockade on the molecule and the sign of the applied bias.   

In Figs.~\ref{fig:prep_tunneling}(a) and \ref{fig:prep_tunneling}(b) we represent the different molecular states prepared by scanning with the tip over the molecule in terms of their pseudospin components. The maps are obtained by simulating an over-threshold bias pulse of $\SI{3}{\ps}$ duration and by computing the pseudospin values corresponding to the end of the pulse. The system is not yet converged to the quasistationary configuration discussed above but shows already clear signatures of the complete blocking state, as can be seen, for example, in the reentrant doublet population in \cref{fig:ev_tunneling} (a). This choice of the pulse duration, will become clearer at a later stage when the role of the SOI will be considered.
The pseudospin shows a prominent dependence on the tip position which is explained by means of \cref{eq:blocking_pseudo} and \cref{eq:tip_pol} in combination with \cref{fig:real_lumos}. 

CuPc has roughly a cross shape. If the tip is placed along the arm horizontal with respect to \cref{fig:prep_tunneling}(a), the bias pulse pumps the pseudospin in the positive $x$ direction. In fact, in the region  $\psi_{yz}^2 > \psi_{xz}^2$, the pseudospin polarization of the tip points in the direction $-{\bm e}_x$ and the pseudospin accumulates in the opposite direction. In other terms, the orbital $L_{xz}$ will be mainly occupied. Vice versa,  the blocking state has mainly a $L_{yz}$ component when the tip moves along the vertical arm of CuPc. Sizable oscillations in the $y$ component of the pseudospin, as seen in Figs.~\ref{fig:prep_tunneling}(b) and \ref{fig:prep_tunneling}(d), show moreover that a linear superposition of both orbitals is required to block the current. Almost no pseudospin polarization is achieved along the horizontal and vertical nodal planes, despite also here one of the two orbitals dominates over the other. The tunneling rate is though so small that the interference blocking mechanism cannot be visible within the \SI{3}{ps} pulse time and the configuration of the molecule remains much closer to the initial thermal state, in which $\langle \boldsymbol{\tau} \rangle = 0$.  

\subsection{Influence of the spin-orbit interaction}

The effect of the SOI on the spectrum analyzed in Sec.~\ref{sec:model} is captured by an effective Hamiltonian, with the same block-diagonal structure as that one of the density matrix discussed so far, since the SOI conserves particle number and cannot split the time-reversal-symmetric singlet states. The doublet and singlet states experience just a slight shift in energy due to the SOI but are not mixed with any other state. Thus, the corresponding blocks are proportional to the identity matrix and not of interest for the dynamics of the molecule.
The block acting on the triplet states is the most interesting part of the SOI effective Hamiltonian. In the basis 
$\ket{T^{+1}_+}$, $\ket{T^{+1}_-}$, $\ket{T^{0}_+}$, $\ket{T^{0}_-}$, $\ket{T^{-1}_+}$, $\ket{T^{-1}_-}$ it reads  
\begin{equation}
    \hamil[SOI]^\mathrm{t} = \begin{pmatrix}
        \nicefrac{\alpha_1}{2}&0&0&0&0&\alpha_3\\
        0&-\nicefrac{\alpha_1}{2}&0&0&\alpha_2&0\\
    0&0&\alpha_4&0&0&0\\
    0&0&0&\alpha_4&0&0\\
    0&\alpha_2&0&0&-\nicefrac{\alpha_1}{2}&0\\
    \alpha_3&0&0&0&0&\nicefrac{\alpha_1}{2}
    \end{pmatrix} + \alpha_5\,\mathbb{1}_6.
\end{equation}
with $\alpha_1 = \SI{0.86}{\meV}$, $\alpha_2 = \SI{1.7e-2}{\meV}$, $\alpha_3 = \SI{8.27e-2}{\meV}$,
$\alpha_4 = \SI{1.0e-2}{\meV}$ and $\alpha_5=\SI{-7.3 e-2}{\meV}$.\\
The projection of $\hamil[SOI]^{\rm t}$ on the operator basis $\mathcal{B}^{\mathrm{t}}$,
\begin{equation}
\begin{aligned}
    \hamil[SOI]^\mathrm{t} &= (\alpha_4+\alpha_5)\mathbb{1}_6+\alpha_1 S_z\otimes\tfrac{\sigma_z}{2}-\alpha_4 S_{z^2}\otimes\mathbb{1}_2\\
    &+(\alpha_2+\alpha_3) S_{x^2y^2}\otimes\tfrac{\sigma_x}{2} + (\alpha_2-\alpha_3)S_{xy}\otimes\tfrac{\sigma_y}{2},
\end{aligned}
\end{equation}
highlights its ability to generate entangled evolution of spin and pseudospin, due to the presence of several product operator components which do not allow us to factorize $\hamil[SOI]^\mathrm{t}$ as the tensor product of a spin and a pseudospin operator. We show in Figs.~\ref{fig:prep_tunneling}(c) and \ref{fig:prep_tunneling}(d) the preparation of the pseudospin in presence of the SOI. Qualitatively, we obtain the same texture of positive and negative areas for the $x$ and $y$ component of the pseudospin, but characterized by an overall suppression of the absolute values.  To analyze this numerical result, we refer once more to the equation of motion for expectation values of the pseudospin.  

The only component of the $\hamil[SOI]$ able to generate an appreciable pseudospin dynamics during the $\SI{3}{\ps}$ time lapse of the bias pulse considered so far is the one proportional to $\alpha_1$. The latter reads $\alpha_1 S_z\otimes\tfrac{\sigma_z}{2}$ and is factorized into the tensor product of a spin and a pseudospin operator. It is thus convenient to study the dynamics of the correlators
\begin{equation}
\begin{split}
  \langle \boldsymbol{\tau}^{\pm} \rangle &:= 
  \Tr[]{\rho^{\rm t}
  \left[
  \left(\frac{1}{3}\mathbb{1}_3 +\frac{1}{6}S_{z^2} \pm \frac{1}{2}S_z\right)\otimes\frac{\boldsymbol{\sigma}}{2}
  \right]},\\
  \langle \boldsymbol{\tau}^0 \rangle &:= 
  \Tr[]{\rho^{\rm t}
  \left[
  \frac{1}{3}(\mathbb{1}_3 - S_{z^2})\otimes \frac{\boldsymbol{\sigma}}{2}
  \right]},
\end{split}
\end{equation}
which are the projections of the pseudospin operator on the subspaces with spin component $S_z = \pm 1$ and $S_z = 0$, respectively. Their equations of motion follow from the generalized master equation $\dot{\rho} = (\liou[M] + \liou[T]) \rho$, approximated to retain only the contribution of the SOI Hamiltonian generating the fastest dynamics  
\begin{figure}
    \centering
    \includegraphics[width=\linewidth]{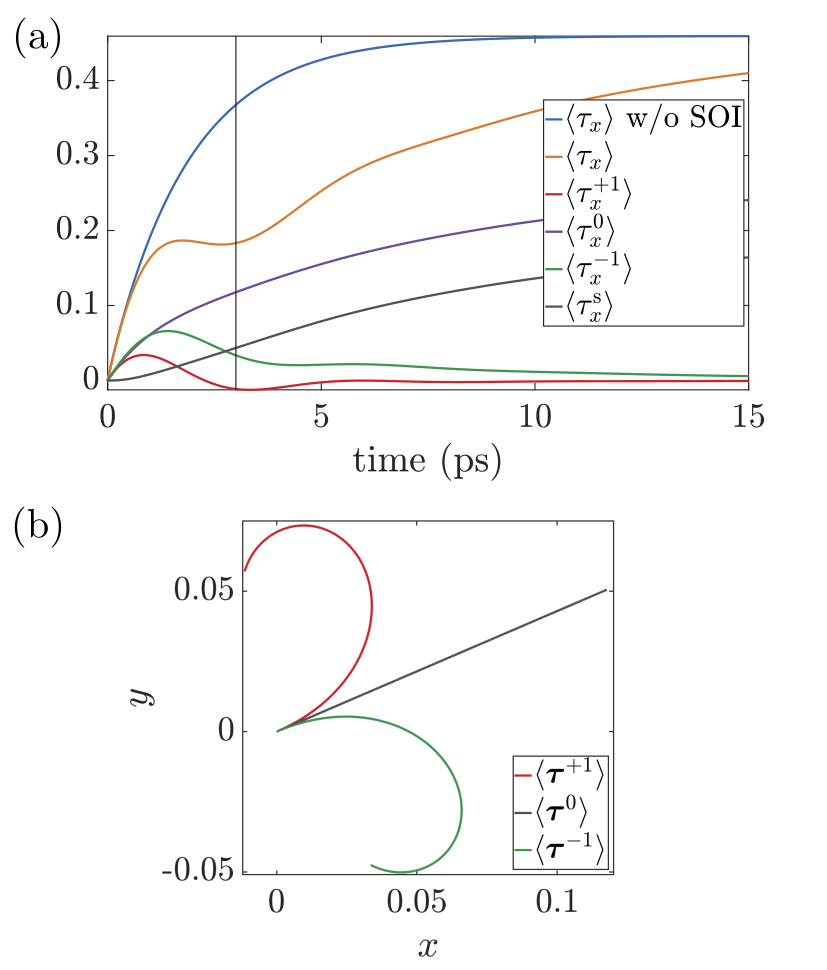}
    \caption{Dynamics of the $x$ components of the pseudospin in constant high bias. (a) Comparison between the total pseudospin $\langle \tau_x \rangle$ and its singlet $\langle \tau^{\rm s}_x \rangle$ and spin-resolved triplet components $\langle \tau^{+1}_x \rangle$, $\langle \tau^{0}_x \rangle$, and $\langle \tau^{-1}_x \rangle$. The SOI is responsible for the shoulder appearing in the time evolution of $\langle \tau_x \rangle$ around $t=\SI{3}{ps}$ and for the oscillations in the $S_z = +1$ and $S_z = -1$ components. (b) Trajectories corresponding to the first $\SI{3}{ps}$ evolution of the spin resolved components of the pseudospin. The vectors $\langle \boldsymbol{\tau}^{\pm 1} \rangle$ undergo precession dynamics inverse to each other due to the SOI, superimposed to the pumping along the direction opposite to ${\bm P}_\tau$. 
    The $\ev{\bm{\tau}^0}$ component is not influenced by the SOI and just grows along the direction opposite to $\bm{P}_\tau$.}
    \label{fig:soi_evolution}
\end{figure}
\begin{equation}
\label{eq:EOM_pseudospin_wSOI}
    \begin{aligned}
        \frac{d}{dt} \ev{\bm{\tau}^{\pm 1}} &=  -\sum_\eta \gez f^-_\eta(\epsilon_\mathrm{t})
         \ev{\bm{\tau}^{\pm 1}} \pm \frac{\alpha_1}{\hbar} \bm{e}_z \times \ev{\bm{\tau}^{\pm 1}}\\
        &+ \gtz \left(\frac{1}{2}\fpt p^{{\rm d}} -\frac{1}{2}\fmt p^{\pm 1}\right) \bm{P}_\tau,\\
        \frac{d}{dt}\ev{\bm{\tau}^0} &=-\sum_\eta \gez f^-_\eta(\epsilon_\mathrm{t})
         \ev{\bm{\tau}^{0}} \\
        &+ \gtz\left(\frac{1}{2} \fpt p^\mathrm{d}-\frac{1}{2}\fmt p^{0}\right) \bm{P}_\tau,
    \end{aligned}
\end{equation}
where $p^{\pm1}$ and $p^{0}$ are the probability to find the system in the triplet states with $S_z=\pm1$ and $S_z = 0$, respectively. The SOI acquires, in the subspace with $S_z = \pm 1$, the form of a pseudomagnetic field pointing along the $z$ direction. Consequently, the dynamics generated by the tunneling Liouvillean, common to the subspace with $S_z = 0$, is enriched here by a precession around the $z$ axis, with opposite orientation for the different spin sectors. 

The time evolution of the different components of the pseudospin under a constant high bias condition can be seen in \cref{fig:soi_evolution}(a). The component of the total pseudospin pointing upwards or downwards along the $x$ direction shows, in presence of SOI, a characteristic shoulder around $t=\SI{3}{ps}$ whose origin is clearly revealed by the observation of the spin resolved components. Both $\langle \tau^\pm_x \rangle$ turn down and more than compensate the constant increase of the  $\langle \tau^0_x \rangle$ component. The deviation from the blocking direction, i.e., the one antiparallel to the pseudospin polarization ${\bm P}_\tau$, characterizing the $S_z = \pm1$ components of the pseudospin is shown in \cref{fig:soi_evolution} (b). The SOI-induced precession allows for these spin states the tunneling coupling at the tip, which is instead soon forbidden, due to the direction of their pseudospin components, to the $S_z=0$ triplet and the singlet states. The latter turn into probability sinks and, on the typical time-scale of a few picoseconds, absorb all the probability and drive the system into a dark state. 

On the intermediate time-scales the SOI is thus responsible for a precession dynamics which reduces the component of the pseudospin along the blocking direction, thus weakening the pseudospin pattern obtained as a function of the tip position after a pump pulse, see \cref{fig:prep_tunneling}. Populating linear superpositions of the states $\ket{T^{+1}_\pm}$ and $\ket{T^{-1}_\pm}$ is moreover fundamental to achieve SOI-induced coherent evolution between the pump and the subsequent probe pulse, as will be discussed  in Sec.~\ref{sec:evolution}.

\subsection{Lamb shift contribution to the preparation}
The last step in the analysis of the complex spin and pseudospin coupled dynamics characterizing the CuPc during the pump pulse concerns the Lamb shift correction to the coherent evolution, introduced in \cref{eq:L_LS} in Sec.~\ref{sec:transport}. The Lamb shift correction stems from electronic fluctuations between the molecule and the leads which induce virtual transitions of the molecule between states with neighboring particle number. Contrary to the tunneling dynamics, though, the Lamb shift correction only arises in systems with level degeneracies and conserves the particle number on the system. In the singular coupling limit  $\liou[LS]\rho = -\tfrac{\iu}{\hbar}\comm{\hamil[LS]}{\rho}$ with the Lamb shift Hamiltonian defined by
\begin{widetext}
\begin{equation}
\label{eq:H_LS_gen}
    \begin{aligned}
        \hamil[LS] = - \frac{1}{2\pi} \sum_{NE\eta} \sum_{i j \sigma}  \Gamma^{\eta}_{i\sigma j\sigma} \proj[NE] \Big[ d^\dagger_{i\sigma} p_\eta(E - \hamil[0]) d_{j\sigma}
        + d_{j\sigma} p_\eta(\hamil[0]-E) d^\dagger_{i\sigma} \Big]\proj[NE],
    \end{aligned}
\end{equation}
\end{widetext}
where $\proj[NE] = \sum_m \ket{NEm}\!\!\bra{NEm}$ is the projection operator on the (degenerate) energy level with particle number $N$ and energy $E$. The collective index $m$ labels the different degenerate energy levels. Due to the singular coupling limit, we neglect the SOI  and only $\hamil[0]$ is taken into account in the calculation of $\hamil[LS]$.    
The logarithmic tails of the digamma function, force us to keep the entire spectrum of both neighboring particle numbers into account. We thus avoid convergence problems in the calculation of $\hamil[LS]$. For the case at hand the contributions of the high-energy neutral states compensate, in fact, those of the high-energy doubly charged molecule (CuPc$^{2-}$). 

\begin{figure}
    \centering
    \includegraphics[width=\linewidth]{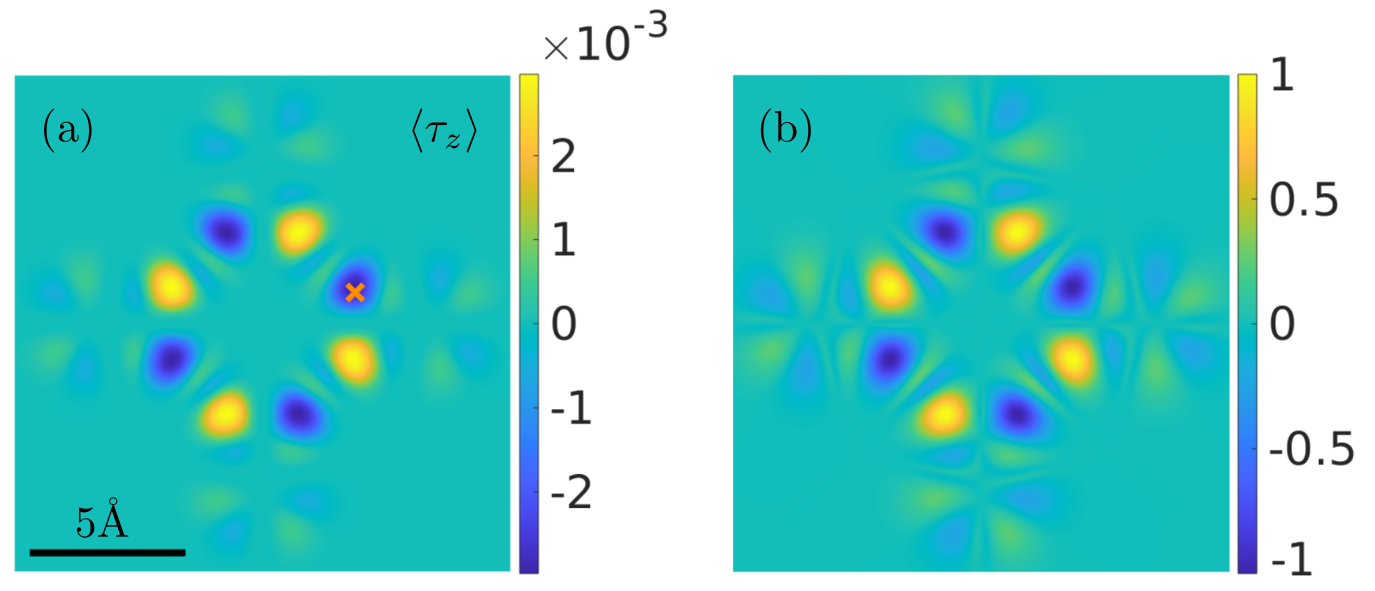}
    \caption{Dependence on the tip position of $\ev{\tau_z}$ after the pump pulse. (a) The combined action of tunneling, SOI, and the Lamb shift during the pump pulse induces nonzero values of $\ev{\tau_z}$. (b) The $z$ component of the torque exerted by the Lamb shift on the pseudospin associated with the dark state normalized to its maximum value, plotted as a function of tip position.}
    \label{fig:tau_z_comparison}
\end{figure}

In \cref{fig:tau_z_comparison}(a) we show the state of the molecule obtained after the pump pulse taking into account the Lamb shift corrections. The Lamb shift correction is responsible of a $\langle \tau_z\rangle$ of the order of $10^{-3}$. In comparison, $\langle \tau_z\rangle$ is negligible, if we only consider the tunneling dynamics and the SOI.  
Despite its modest effects in the pulsed dynamics, the nonzero $\langle \tau_z\rangle$ component obtained here is amplified by the SOI during the free evolution following the pulse (see \cref{fig:full_dynamics}). Moreover, the presence of an imbalance in the occupation of opposite angular-momentum states opens the question about circular currents triggered by the coupling to the leads and their feedback on the electronic structure of the molecule \cite{Schoenauer2019}. 

The results presented in \cref{fig:tau_z_comparison} (a) can be understood by analyzing the structure of the Lamb shift Hamiltonian. The latter is block diagonal in the doublet, triplets and singlets subspaces and spin isotropic. By neglecting the components proportional to the identity, which drop from $\liou[LS]$, we can write the general form for the singlets and triplet blocks, respectively
\begin{equation}
\label{H_LS_pseudoB}
\begin{split}
    \hamil[LS]^{\rm s} &= \hbar\boldsymbol{\omega}^{\rm s}_{\rm LS}\cdot \frac{\boldsymbol{\sigma}}{2}\,,\\
    \hamil[LS]^{\rm t} &= \mathbb{1}_3 \otimes \hbar\boldsymbol{\omega}^{\rm t}_{\rm LS} \cdot \frac{\boldsymbol{\sigma}}{2}\,, 
\end{split}
\end{equation}
from which it is clear that the Lamb shift Hamiltonian acts on the singlet as well as on the triplet subspace as a pseudomagnetic field. The latter adds, during the short pump pulse, to the spin dependent pseudomagnetic field associated with the SOI which was introduced in \cref{eq:EOM_pseudospin_wSOI}. 
The strength and the direction of the pseudo-magnetic fields $\boldsymbol{\omega}^{\rm t}_{\rm LS}$ and $\boldsymbol{\omega}^{\rm s}_{\rm LS}$ are obtained from the direct evaluation of \cref{eq:H_LS_gen} for the anionic subspace of CuPc. By following the derivation in \cite{Sobczyk2012} adapted to the $D_{4h}$ point-group symmetry of the CuPc, \footnote{Particular care should be taken to take into account the different symmetry of the frontier orbitals with respect to the horizontal symmetry plane $\sigma_h$.} we  get
\begin{equation}
    \boldsymbol{\omega}^{\rm s/t}_{\rm LS} = \frac{\bar{\Gamma}^{\rm tip}}{2} 
    (A^{\rm s/t} {\bm P}_\tau + B^{\rm s/t} {\bm Q}_\tau) ,
\end{equation}
where
\begin{equation}
{\bm Q}_\tau = \frac{1}{\psi_{xz}^2 + \psi_{yz}^2}
        \begin{pmatrix}
     \psi_{xz}^2 - \psi_{yz}^2\\
     -2\psi_{xz}\psi_{yz}\\
     0
    \end{pmatrix},
\end{equation}
while $A^{\rm s/t}$ and $B^{\rm s/t}$ are real numbers which depend on the electronic structure of the molecule and on the applied bias across the junction. Specifically, they read, respectively, for the triplet and for the singlet
\begin{equation}
    \begin{split}
        A^{\rm t} =-\frac{1}{6\pi}\sum_{S_z\sigma}
        \Big[
        &\bra{T^{S_z}_+}d^\dagger_{L+,\sigma} p_{\rm T}(E^{\rm t}-\hamil[0])d_{L-,\sigma}\ket{T^{S_z}_-}\\
        + 
        &\bra{T^{S_z}_+}d_{L-,\sigma} p_{\rm T}(\hamil[0]-E^{\rm t})d^\dagger_{L+,\sigma}\ket{T^{S_z}_-}
        \Big],\\
        B^{\rm t} = -\frac{1}{6\pi}\sum_{S_z\sigma}
        \Big[
        &\bra{T^{S_z}_+}d^\dagger_{L-,\sigma} p_{\rm T}(E^{\rm t}-\hamil[0])d_{L+,\sigma}\ket{T^{S_z}_-}\\
        +
        &\bra{T^{S_z}_+}d_{L+,\sigma} p_{\rm T}(\hamil[0]-E^{\rm t})d^\dagger_{L-,\sigma}\ket{T^{S_z}_-}
        \Big],
    \end{split}
\end{equation}
and
\begin{equation}
    \begin{split}
        A^{\rm s} = -\frac{1}{2\pi}\sum_{\sigma}
        \Big[
        &\bra{S_+}d^\dagger_{L+,\sigma} p_{\rm T}(E^{\rm s}-\hamil[0])d_{L-,\sigma}\ket{S_-}\\
        +
        &\bra{S_+}d_{L-,\sigma} p_{\rm T}(\hamil[0]-E^{\rm s})d^\dagger_{L+,\sigma}\ket{S_-}
        \Big],\\
        B^{\rm s} = -\frac{1}{2\pi}\sum_{\sigma}
        \Big[
        &\bra{S_+}d^\dagger_{L-,\sigma} p_{\rm T}(E^{\rm s}-\hamil[0])d_{L+,\sigma}\ket{S_-}\\
        +
        &\bra{S_+}d_{L+,\sigma} p_{\rm T}(\hamil[0]-E^{\rm s})d^\dagger_{L-,\sigma}\ket{S_-}
        \Big].
    \end{split}
\end{equation}
The pseudomagnetic fields acting on the  triplet and singlet subspaces differ, in general, in strength and direction. To achieve a qualitative understanding of the results in \cref{fig:tau_z_comparison} it is useful to concentrate on their dependence on the tip position, expressed through the vectors ${\bm P}_\tau$ and ${\bm Q}_\tau$. Given that the tunneling pumps the pseudospin along the direction $-{\bm P}_\tau$, the direction and frequency of the precession induced by $\hamil[LS]$ at a given bias is readily obtained by the vector product 
\begin{equation}
\label{eq:Torque}
  -\boldsymbol{\omega}_{\rm LS}^{\rm t/s} \times {\bm P}_\tau = -B^{\rm t/s} \gamma_0^{\rm tip}
  \frac{\psi_{xz}\,\psi_{yz} (\psi_{xz}^2 -\psi_{yz}^2)}{\psi_{xz}^2 + \psi_{yz}^2}{\bm e}_z  , 
\end{equation}
which is proportional to the torque acting on the pseudospin when the latter is pumped in the direction of the dark state. Since the time lapse of the pump pulse is less than a quarter of the precession period induced by the Lamb shift, the direction and intensity of the torque are proportional to those of the $z$ component of the pseudospin just after the pulse. In \cref{fig:tau_z_comparison} we see a comparison between the simulated $z$ component of the pseudospin and the torque, normalized to its maximum value, as obtained from \cref{eq:Torque}. 
The torque vanishes when ${\bm P}_\tau$ is parallel or antiparallel to  ${\bm Q}_\tau$. For this reason we see no $z$ component of the pseudospin along the $x$ and $y$ axis where either one or the other LUMO orbital vanishes and along the diagonals $x_{\rm tip} = \pm y_{\rm tip}$ where the two orbitals necessarily assume the same absolute value. Moreover, the torque is strongly suppressed on the outer parts of the ligand arms where one of the two real valued LUMOs has much more weight than the other one. Finally, as can be readily checked from the \cref{eq:Torque}, the pattern of $\langle \tau_z \rangle$ obtained from the pump pulse is invariant under rotations of $\pi/2$ around the main rotational axis and antisymmetric with respect to the four vertical symmetry planes of CuPc. From these properties it derives a certain resemblance to the HOMO of the molecule. 

\section{Free evolution and read out}
\label{sec:evolution}

The intricate dynamics triggered by the pump pulse brings the CuPc molecule from its thermal equilibrium into an excited, coherent superposition of many-body states. The latter, as we have shown, can be controlled by the tip position and the duration of the pump pulse. We focus, in this section, on the following time lapse in which the free evolution in absence of bias reveals an intertwined spin and pseudospin dynamics. Moreover, we introduce a second pulse to achieve a stroboscopic readout and investigate the evolution of the molecule during the delay time between the pump and the probe pulses.     

\subsection{Free evolution}
\label{sec:free_evolution}

\begin{figure}
\begin{center}
    \includegraphics[width=\linewidth]{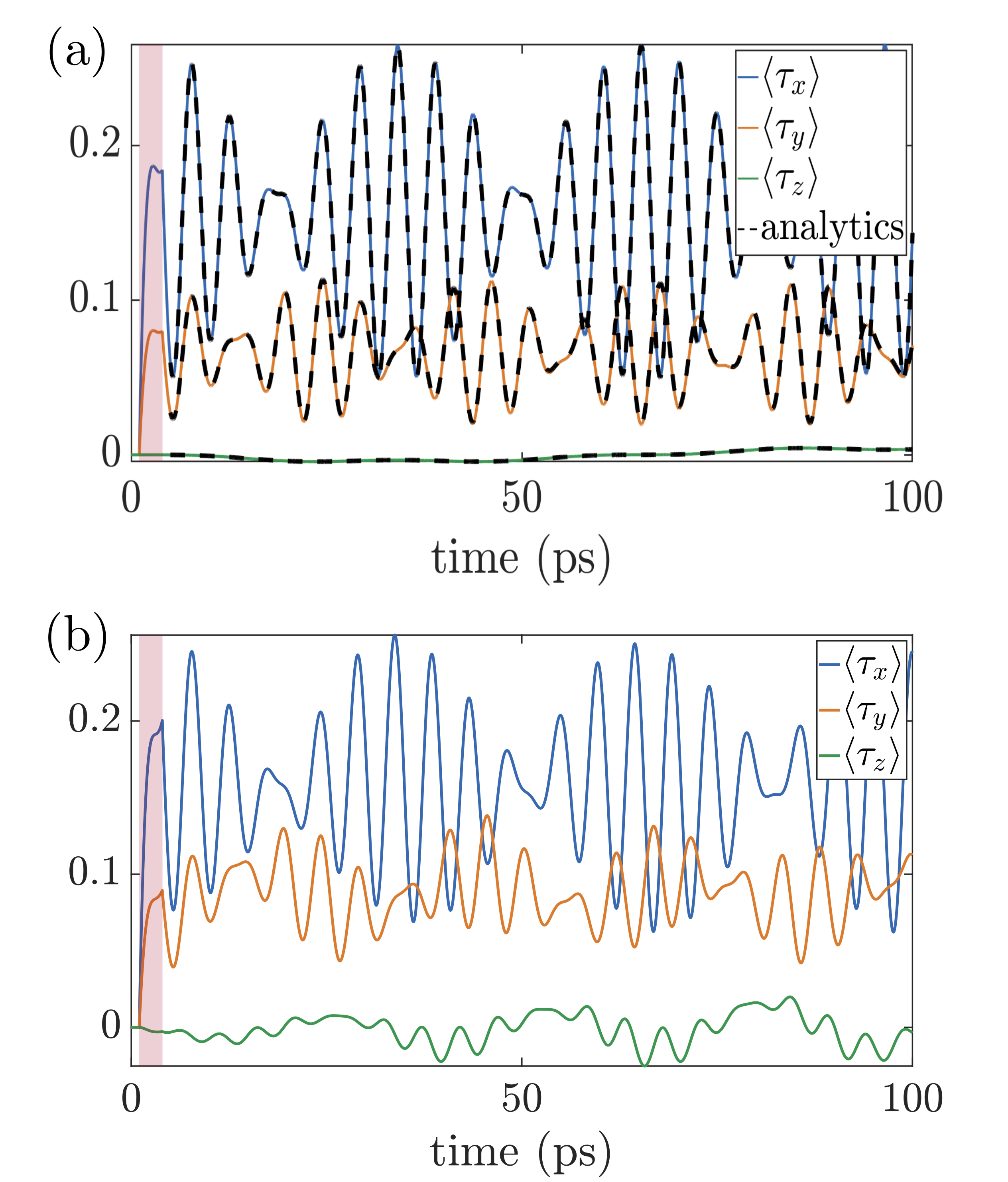}  
\end{center}
\caption{Free evolution of the pseudospin components after the pump pulse. The purple area indicates the 
        time lapse during which the bias pulse is applied. In panel (a) we show the dynamics only driven by the SOI with the analytical solution for the free evolution on top of the numerics. In panel (b) the Lamb shift contribution is also included in the calculation. While $\ev{\tau_x}$ and $\ev{\tau_y}$ show qualitatively the same beating dynamics, $\ev{\tau_z}$ changes radically. The Lamb shift couples in fact more strongly the $\ev{\tau_z}$ dynamics to that of the $x$ and $y$ components, thus mixing the evolution time-scales.}
\label{fig:full_dynamics}
\end{figure}

The state prepared by the pump pulse is a superposition of several anionic states. Once the external bias turns back to zero, tunneling events are strongly suppressed due to the large charging energy of the molecule with respect to temperature and the charge is conserved. The time evolution is described only by the singlet and triplet sub-blocks of the reduced density matrix. Moreover, there is no coupling between the singlet and the triplet states, which thus evolve independently, as an incoherent superposition. Specifically, as the singlet energy eigenstates are connected by time-reversal symmetry, they are degenerate and thus they show no dynamics at all. On the contrary, the triplet states are characterized by an intertwined spin and pseudospin dynamics, due to the SOI, and by the pseudospin precession due to the Lamb shift, as described in the previous section.

In \cref{fig:full_dynamics} we present the time evolution of the three components of the pseudospin during $\SI{100}{ps}$, starting from the beginning of the pump pulse. In \cref{fig:full_dynamics} (a) We first concentrate, for simplicity, on the evolution generated solely by the SOI-Hamiltonian (the contribution of $\hamil[0]$ drops from the Liouvillean, being just a uniform energy shift). The $x$ and $y$ components show beating dynamics with the fast oscillations compatible with the largest SOI-induced splitting $\alpha_1$ and the slow modulation associated with the smaller energy separations $\alpha_2$ and $\alpha_3$ between the two triplet states with the lowest and, respectively, the highest energies. As we already discussed in the previous section, the $z$ component of the pseudospin is conserved by the part of $\hamil[SOI]$ proportional to $\alpha_1$ and thus it can only exhibit slow dynamics. 

Further insight into the free evolution is given by the analytic result plotted as a dashed black line in \cref{fig:full_dynamics}. We obtain it by solving the coherent contribution of the GME $\dot{\rho} = -\tfrac{\iu}{\hbar} \comm{\hamil[SOI]}{\rho}$,
projected into the operator basis introduced in Sec.~\ref{sec:Operator_space}. The $x$ component of the pseudospin evolves as
\begin{equation}
\label{eq:tx_evol}
    \begin{aligned}
        \ev{\tau_x}(t) =& \evz{\tau_x^\mathrm{s}} + \frac{1}{3} \big(\evz{\tau_x^\mathrm{t}}-\evz{S_{z^2}^\mathrm{t}\tau_x^\mathrm{t}}\big) +\\
        \frac{1}{12}\bigg[&\sqrt{a^2+c^2}\cos\left(\omega_{1} t+ \arctan\frac{c}{a}\right) +\\&
        \sqrt{b^2+d^2}\cos\left(\omega_{2} t+\arctan\frac{d}{b}\right)\bigg],
    \end{aligned}
\end{equation}
where the two oscillation frequencies are 
\begin{equation}
\omega_{1} = \frac{1}{\hbar}(\alpha_1+\alpha_2-\alpha_3)\quad  \text{and} \quad
\omega_{2} = \frac{1}{\hbar}(\alpha_1-\alpha_2+\alpha_3),
\end{equation}
while the amplitudes and phases are expressed in terms of the four coefficients
\begin{equation}
\begin{split}
a &= 4\evz{\tau_x^\mathrm{t}}+2\evz{S_{z^2}^\mathrm{t}\tau_x^\mathrm{t}}-3\evz{S_{x^2y^2}^\mathrm{t}},\\ b &= 4\evz{\tau_x^\mathrm{t}}+2\evz{S_{z^2}^\mathrm{t}\tau_x^\mathrm{t}}+3\evz{S_{x^2y^2}^\mathrm{t}},\\ c &= 6(\evz{S_z^\mathrm{t}\tau_y^\mathrm{t}}-\evz{S_{xy}^\mathrm{t}\tau_z^\mathrm{t}}),\\ 
d &= 6(\evz{S_z^\mathrm{t}\tau_y^\mathrm{t}}+\evz{S_{xy}^\mathrm{t}\tau_z^\mathrm{t}}),
\end{split}
\end{equation}
which contain several triplet correlators calculated at the time $t=0$ at which the pump pulse terminates. 
In the first line in \cref{eq:tx_evol} one recognizes the constant terms associated with the singlet and the $S_z = 0$ component of the pseudospin. The second and the third line, instead, describe the dynamics of the $S_z = \pm1$ components intertwined by the SOI. The close frequencies $\omega_1$ and $\omega_2$ are at the origin of the beatings observed in \cref{fig:full_dynamics}.
Moreover, it is clear that the pseudospin at time $t > 0$ depends on several mixed correlators,  e.g.$\evz{S_{z^2}^\mathrm{t}\tau_x^\mathrm{t}}$ calculated at initial time, thus revealing the interplay of spin and pseudospin in the evolution of CuPc. 

The analogous expressions for the time evolution of the $y$ and $z$ components of the pseudospin can be found in the Sec.~\ref{sec:tytz_evolution}. Here we just mention that the $y$ component combines oscillations with frequencies $\omega_3 = (\alpha_1 - \alpha_2 - \alpha_3)/\hbar$ and $\omega_4 = (\alpha_1 + \alpha_2 + \alpha_3)/\hbar$, thus featuring also beating dynamics, while the $z$ component contains only the smaller frequencies $\alpha_2/\hbar$ and $\alpha_3/\hbar$. In summary, all possible Bohr frequencies associated with the two lowest and two highest triplet states are reflected into the dynamics of the pseudospin.  

\begin{figure}
     \centering
     \includegraphics[width=\linewidth]{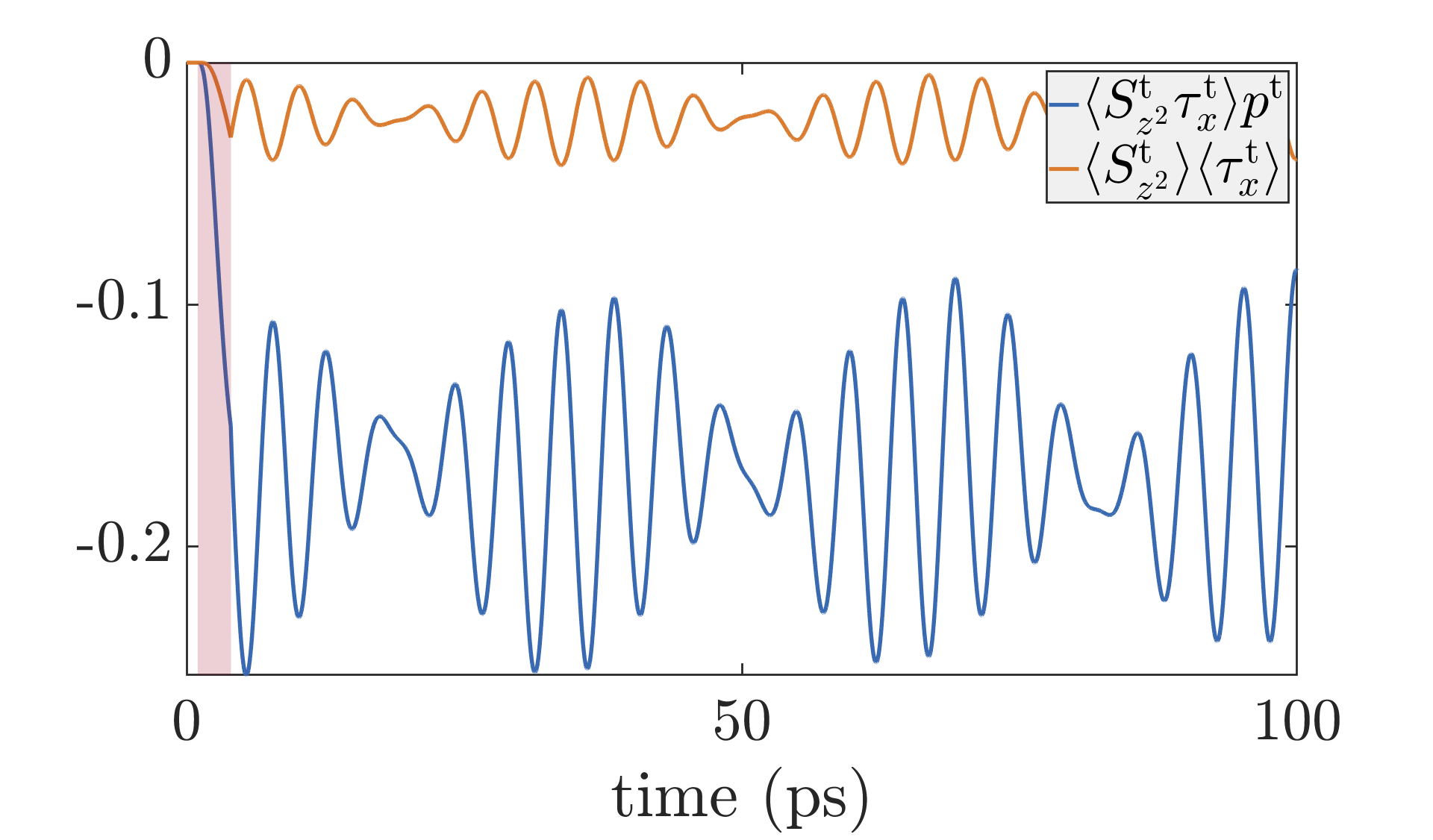}
     \caption{Dynamics of a mixed correlator. Also the mixed correlators, such as, for example, $\langle S_{z^2}^{\rm t}\tau_x^{\rm t}\rangle$, show beatings dynamics. Moreover, the comparison with product of the single spin and pseudospin expectation values reveals the correlation between these two degrees of freedom triggered by the SOI.}
     \label{fig:evolution_corr}
 \end{figure}

The spin and orbital correlation is further demonstrated by the comparison of the mixed triplet correlators with the product of the single expectation values. As shown in \cref{fig:evolution_corr} the mixed correlators are essentially equal to their factorized counterpart at $t=0$ when the reduced density matrix is in thermal equilibrium. The situation changes drastically during the pump pulse and remains essentially the same during the free evolution, thus demonstrating the correlation between the two degrees of freedom of the molecule. 

Eventually, we show in \cref{fig:full_dynamics} (b) the combined effects of the SOI and the Lamb shift correction on the evolution of the CuPc anionic state. The combination of the effective magnetic fields associated with the SOI and to the the Lamb shift correction, the first oriented along the $z$ axis, the second in the $xy$ plane, mixes the time evolution of the different pseudospin components. The result is a more complex, but also more homogeneous oscillation beating pattern, now visible in all three components.  

\subsection{Transferred charge}

In this section we propose a readout scheme for the observation of the SOI-induced dynamics described so far. To this end we simulate a short ($\SI{0.5}{ps}$) over-threshold probe pulse following the pump pulse at a variable delay time. The observable is the total charge transferred through the molecule and collected into the tip, calculated as a function of the delay time. 

\begin{figure}
    \centering
    \includegraphics[width=\linewidth]{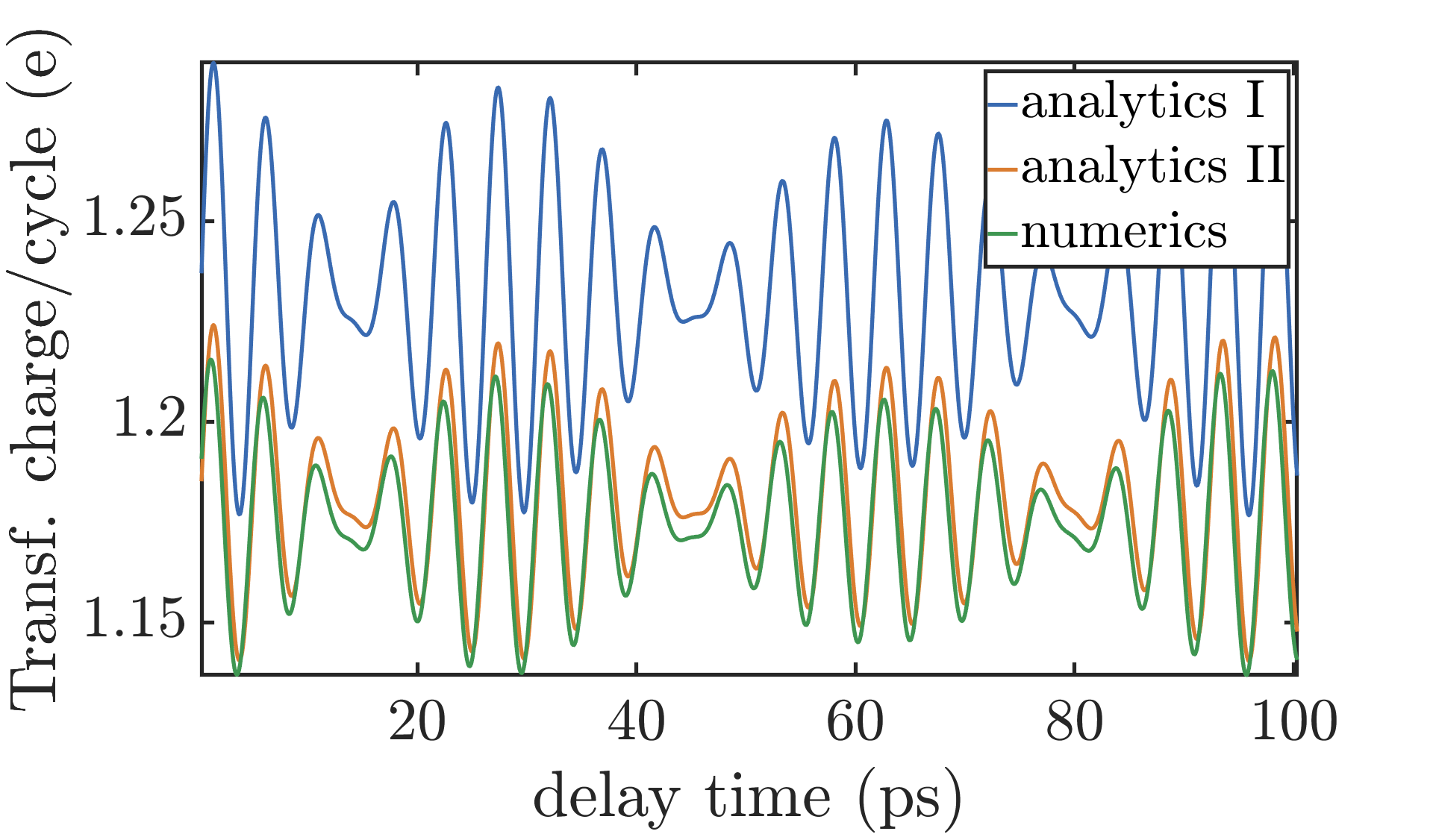}
    \caption{Transferred charge per pump probe cycle plotted as a function of the delay time between the two pulses. 
    The blue and the orange solid lines represent analytical results calculated according to \cref{eq:transf_charge} and its improved version, \cref{eq:transf_charge_improved}, respectively.}
    \label{fig:charge_trans}
\end{figure}

In \cref{fig:charge_trans} we show the result of our simulations, with the charge transferred to the tip  per pump probe cycle in units of the electronic elementary charge. Despite the sequential tunneling regime, fractional charge can be transferred through the molecule. The result should be interpreted in a probabilistic sense, with the charge averaged over a great number of pump probe cycles. The high-frequency repetition of the pump probe pulse is anyway necessary in the experiments, with the aim to obtain a measurable average current, i.e., the mean transferred charge per pump probe cycle divided by the repetition period \cite{Cocker2016,Peller2020}.
It is crucial, though, for the meaningful comparison of the experimental data with the theoretical predictions presented here that the repetition period be longer than the relaxation time of the molecule towards its thermal equilibrium. The latter is in fact taken in our simulations as the starting point of the excitations induced by each pump pulse.

The beatings in the transferred charge per cycle presented in \cref{fig:charge_trans} closely resemble those obtained in the free evolution of the pseudospin shown in \cref{fig:full_dynamics} (b), thus indicating that the proposed measurement of the transferred charge can give information about the SOI-induced dynamics on the system. 

This similarity is rationalized starting from the following observations: The dynamics induced on the junction by the probe pulse closely resembles the one produced by the pump pulse, with the bias across the junction allowing energetically the transition between the anionic and the neutral states of the  molecule. There are, though, two crucial differences.  First, the probe pulse impinges on the molecule when the latter is far from equilibrium: it is instead characterized by a significant population of the singlet states, a finite pseudospin, an anisotropic spin state and it even shows correlations between the spin and the pseudospin degrees of freedom, as we demonstrated in the previous sections. Second, due to its shorter duration, the probe pulse does not induce a large variation of the pseudospin on the system, although it is long enough to ensure, in principle, a measurable charge transfer towards the tip. 

Crucially, the probability for charge transfer to occur during the probe pulse depends on the state of CuPc at the beginning of the pulse. The closer is the molecular state to the electronic dark state [see the condition given in \cref{eq:blocking_pseudo}], the less probable is the tunneling event which brings an electron from the molecule towards the tip. Because the pseudospin dynamics during the delay time between the pump and the probe pulses is strongly influenced by the SOI, the latter leaves its clear fingerprints also in the average transferred charge. 

In a first approximation, we can formalize the ideas just expressed in the following equation for the transferred charge:
\begin{equation}
\label{eq:transf_charge}
    q_{\rm tr}(t) = q_0 + \bar{\Gamma}^{\rm tip} \delta t 
    \left[1 + 2 \langle \boldsymbol{\tau} \rangle (t) \cdot {\bm P}_\tau\right],
\end{equation}
where $q_0$ is the charge transferred during the pump pulse, $\delta t$ is the length of the probe pulse and $\langle \boldsymbol{\tau} \rangle (t)$ is the pseudospin of the molecule at the delay time $t$, i.e., at the beginning of the probe pulse.
This result is derived from the formal expression for the current given in \cref{eq:current} under the approximation that, during the probe pulse, neither the jump operators nor the reduced density matrix for the system depend on time and in the high bias limit in which unidirectional transport is enforced. 

\begin{figure}
    \centering
    \includegraphics[width=\linewidth]{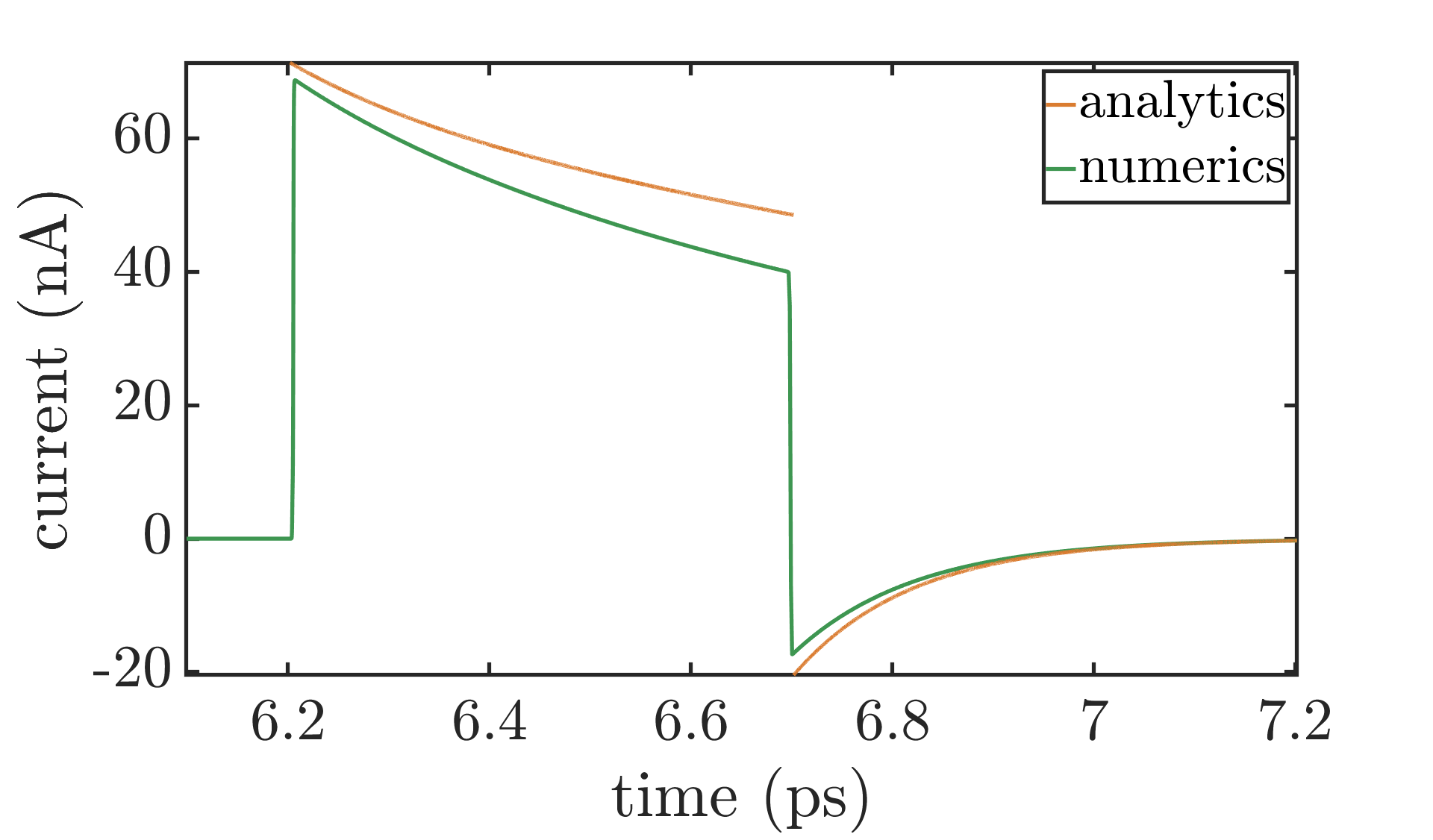}
    \caption{Current flowing towards the tip during the probe pulse. The current decay during the probe pulse and the back flow from the tip to the molecule are essential for a quantitative reconstruction of the readout signal.}
    \label{fig:curr_tip}
\end{figure}

Although the essence of the readout mechanism is captured by \cref{eq:transf_charge}, the latter quite strongly overestimates the charge transfer during the probe pulse, as seen in \cref{fig:charge_trans}. Two are the fundamental reasons of this quantitative mismatch: the system evolution during the $\SI{0.5}{ps}$ probe pulse and the back flow of electrons from the tip to the molecule, happening shortly after the end of the probe pulse. Both effects are clearly visible in the current simulation performed for times in the vicinity of the probe pulse and are shown in \cref{fig:curr_tip}. 

A substantially improved, although more intricate, expression for the transferred charge is given by the integrals 

\begin{equation}
\label{eq:transf_charge_improved}
\begin{split}
    q_{\rm tr}(t) = q_0 &+ 
    \bar{\Gamma}^{\rm tip}\int_{t}^{t+\delta t} \!\!
     \left[1 - p^{\rm d}(t') + 2 \langle \boldsymbol{\tau} \rangle (t') \cdot {\bm P}_\tau \right]  {\rm d}t' \\
     &- \bar{\Gamma}^{\rm tip}\int_{t+\delta t}^\infty \!\! 4 p^{\rm d}(t') {\rm d}t',
     \end{split}
\end{equation}
derived as was \cref{eq:transf_charge} by assuming unidirectional transport during the probe pulse, but including at the same time the system dynamics. Moreover, the back flow after the probe pulse is accounted for. The latter is due to the relaxation of the molecule towards its anionic configuration, occurring when the bias drops to zero, at the end of the probe pulse. At first, it seems detrimental for the efficiency of the readout protocol that the correct estimate of the transferred charge would require us to monitor the current for an infinitely long time. De facto, the complete decay of the doublet population is reached within a few hundred femtoseconds, thus reaffirming the  feasibility of the proposed protocol.

The transferred charge calculated according to \cref{eq:transf_charge_improved} is shown in \cref{fig:charge_trans} and matches much better the benchmark numerics, as compared with \cref{eq:transf_charge}. The corresponding analytical calculation of the time-dependent current is plotted in \cref{fig:curr_tip} together with the corresponding numerical simulation. 

We calculate the transferred charge \cref{eq:transf_charge_improved} analytically, only accounting for the tunneling dynamics. The set of coupled differential equations for $p^{\rm d}$ and $\langle\boldsymbol{\tau}\rangle$ derived from $\dot{\rho} = \liou[T] \rho$ is solved with initial conditions $p^{\rm d} = 0$ and $\langle\boldsymbol{\tau}\rangle$ extracted from the numerical simulation at the beginning of the probe pulse. Eventually, the transferred charge can be cast in the same form as in \cref{eq:transf_charge}, with just the probe time lapse $\delta t$ being renormalized by the factor
\begin{equation}
\begin{split}
    K &= 
    \left( 1 - \frac{1 - R}{\sqrt{1+R^2}} \right)\frac{1}{2\lambda_1 \delta t}
    \left(1 - e^{-\lambda_1 \delta t}  \right)\\
    & +
    \left( 1 + \frac{1 - R}{\sqrt{1+R^2}} \right)\frac{1}{2\lambda_2 \delta t}
    \left(1 - e^{-\lambda_2 \delta t} \right)\\
    & -
    \frac{1}{(2+R)\sqrt{1+R^2}} \frac{1}{\bar{\Gamma}^{\rm tip}\delta t }
    (e^{-\lambda_1 \delta t} - e^{-\lambda_2 \delta t}),
    \end{split}
\end{equation}
where $R = 2 \bar{\Gamma}^{\rm sub}/\bar{\Gamma}^{\rm tip}$, while the rates $\lambda_{1,2}$ read
\begin{equation}
\lambda_{1,2} = \bar{\Gamma}^{\rm tip}(1+R \mp \sqrt{1+R^2}).
\end{equation}
The small mismatch between the analytics and the numerical simulations which can be seen in Figs.~\ref{fig:charge_trans} and \ref{fig:curr_tip} is due to the dynamics induced by the SOI during the probe pulse, still neglected for simplicity in the the evaluation of \cref{eq:transf_charge_improved}.

Interestingly, the renormalization factor $K$ only depends on the two dimensionless parameters $R$ and $\bar{\Gamma}^{\rm tip}\delta t$, which largely determines the physics of the problem. For example, for pulses much shorter than the time-scales set by the tip tunneling rate ($\bar{\Gamma}^{\rm tip}\delta t \ll 1$) we can greatly simplify the renormalization factor to obtain
\begin{equation}
    \lim_{\bar{\Gamma}^{\rm tip}\delta t \to 0} K = \frac{\bar{\Gamma}^{\rm sub}}{\bar{\Gamma}^{\rm sub} + \bar{\Gamma}^{\rm tip}}.
 \end{equation}

If, moreover, we consider $\bar{\Gamma}^{\rm sub} \gg \bar{\Gamma}^{\rm tip}$, the result in \cref{eq:transf_charge} is obtained again. The physical interpretation of this limit is quite clear: the dynamics of the molecule during the pump pulse can be to a large extent neglected and moreover the recharging of the molecule after the probe pulse happens mainly at the substrate side, thus hindering the back flow from the tip towards the molecule. 

In the opposite limit, $\bar{\Gamma}^{\rm sub} \ll \bar{\Gamma}^{\rm tip}$, a very small signal is expected in the read out, since in this case the back flow would almost completely balance the current flowing from the molecule towards the tip during the probe pulse.

Finally, in the limit of long probe pulses $\bar{\Gamma}^{\rm tip}\delta t \gg 1$, the charge transferred during the probe pulse converges to the value of $1 + 2 \langle \boldsymbol{\tau}\rangle (t)\cdot {\bm P}_\tau$ and is independent of the tunneling rates. 

The information about the SOI-induced dynamics is included in \cref{eq:transf_charge} (and in its improved version with the renormalized factor $K$) through the pseudospin $\langle \boldsymbol{\tau} \rangle (t)$ calculated at the delay time $t$. Indeed, as we have already discussed, the pseudospin beatings contain all Bohr frequencies of the triplet low-energy states. Thus, monitoring the dynamics of the transferred charge gives access to the SOI-induced spectrum. Moreover, an interesting aspect of the molecular many-body state can be observed, since we can also read out the projection of the pseudospin along the known direction ${\bm P}_\tau$. 

Eventually, we expect the transferred charge to be also strongly dependent on the tip position. On the one hand, the overall constant shift $q_0$ depends on the position as well as the average tip rate which is proportional to $\psi_{xz}^2({\bm r}_{\rm tip}) + \psi_{yz}^2({\bm r}_{\rm tip})$. Moreover, the tip position influences the prepared pseudospin since both the dark state as well as the torque induced by the Lamb shift show a characteristic pattern. The tip position can even modulate the very dynamics of the pseudospin via the pseudomagnetic fields $\boldsymbol{\omega}^{\rm s/t}_{\rm LS}$. The analysis of these effects, though, goes beyond the scope of this work.   

\section{Conclusions}
\label{sec:conclusions}

The theoretical investigation of a lightwave-STM single molecule junction is on the focus of this paper. In particular, we studied how to access, via a stroboscopic pump probe scheme, the spin-orbit-induced electronic dynamics of a copper phthalocyanine. To this end, we extended the transport theory for STM on thin insulating films \cite{Sobczyk2012} to include the effects of laser pulses, specific of a lightwave-STM, directly in the time domain. We identify in the coherent superposition of different angular-momentum states the fundamental mechanism for the triggering and read out of spin-orbit-induced dynamics. The method of choice is a generalized master equation (GME) for the reduced density matrix, which allows for the treatment of strong correlations on the CuPc weakly coupled to both the substrate and the tip. The tunneling and the SOI-induced dynamics share similar time-scales resulting in a nonadiabatic time evolution of the system.   

The localized tunneling at the tip combined with the quasidegenerate spectrum of the molecule allows for interference effects to appear. The pump pulse takes the molecule out of its thermal equilibrium and brings it towards an electronic dark state, associated with a specific pseudospin configuration, i.e., a coherent superposition of angular-momentum states. After the lightwave-induced bias is turned off again, the system shows an intricate beating dynamics of the pseudospin.   The latter is driven at the same time by the spin-orbit interaction and by virtual electronic fluctuations between the molecule and the tip (the Lamb shift). In particular, oscillations at all Bohr frequencies associated with the spectrum of $\hamil[SOI]$ contribute to the time evolution of the pseudospin. 

Furthermore, we propose to use a second, shorter, bias pulse in order to probe the many-body state acquired by the system after a given delay time separating the two pulses. Specifically, the charge transferred during the probe pulse is proportional to the projection of the pseudospin along the known dark-state direction. Ultimately, monitoring the beatings in the average charge per pump probe cycle allows one to access the spin-orbit-induced dynamics of a CuPc. 

Both the electronic dark state as well as the pseudospin torque induced by the Lamb shift vary with tip position with patterns obtained by different combinations of the LUMO wave functions, as can be seen by comparing Figs. \ref{fig:prep_tunneling} and \ref{fig:tau_z_comparison} with Eqs. \eqref{eq:tip_pol} and \eqref{eq:Torque}. We expect those patterns to leave their signatures in the pseudospin dynamics and ultimately to emerge in the transferred charged recorded as a function of tip position and time. The high temporal and spatial resolution demonstrated in lightwave-STM experiments \cite{Cocker2016,Peller2020} is certainly capable to detect those patterns. 

Lightwave-STM is a powerful, novel technique, which allows us to access the molecular dynamics directly on the intrinsic atomic-spatial and femtosecond time-scales. The control and read out of molecular quantum states proposed in this paper, suggest its application also for molecular-based quantum information, a novel and challenging development in the field of molecular electronics. 

Moreover, we expect the theoretical method presented in this paper to open new paths for the investigation of proximity effects in molecular junctions. Exemplary, a ferromagnetic tip or substrate could be considered. Besides working as sources of polarized electrons, they can induce proximity effects in single-molecule junctions. Dipolar (and also quadrupolar for systems with high spin, $S > \tfrac{1}{2}$) exchange fields arise, which influence the electronic dynamics on the molecule and its transport characteristics. Such proximity effects could be simultaneously controlled and monitored in a spin-polarized lightwave-STM.

Finally, with the same investigation tool, the interplay of electrical and mechanical degrees of freedom on the molecular level could also be investigated. Thus, along the lines described in this article, simulation of lightwave-STM could contribute to the understanding of surface chemical reactions,  followed directly in real time.

\section{Acknowledgments}
We thank Milena Grifoni and Jascha Repp for fruitful discussions. We acknowledge moreover financial support from the DFG via SFB-1277 (Subproject B02).

\appendix
\section{Quasi-Stationary High-Bias Limit}
\label{sec:tunneling}

The molecule does not reach its steady state during the $\SI{3}{ps}$ pump pulse. Although the eigenvalues of the reduced density matrix as well as the pseudospin components are apparently converged (see \cref{fig:prep_tunneling}), also the state reached after the $\SI{15}{ps}$ full simulation time is also not the steady state. The latter, despite the high-bias condition, should in fact display a Boltzmann ratio  between the population of the singlet and the triplet states, as explained in the main text. 
To explain this result we consider the bare tunneling dynamics expressed in the basis of the coupled and decoupled states, i.e.,  the one containing, for the singlet and triplet subspaces, the dark state (decoupled) complemented by its orthogonal companion. Since we do not consider SOI nor the Lamb shift contribution, the coherences vanish , in the basis just described, and we can concentrate on the populations only.
The basis reads then 
\begin{equation}
    \label{eq:basis_cd}
    \begin{aligned}
    &\liouket{D^\uparrow},\liouket{D^\downarrow},\liouket{T^{+1}_c},\liouket{T^{+1}_d},\liouket{T^{0}_c},
    \liouket{T^{0}_d},\\ & \liouket{T^{-1}_c},\liouket{T^{-1}_d},\liouket{S_c},\liouket{S_d}.
    \end{aligned}
\end{equation}
where $\liouket{\cdot}$ indicates states in Liouville space.

The tunneling Liouvillean written in the basis (\ref{eq:basis_cd}) has the following form
\begin{equation}
    \begin{aligned}
    &\liou[tun] =\\& \begin{pmatrix}
        \liou[D^\uparrow D^\uparrow] & 0 & \liou[D^\uparrow T^{+1}] & \liou[D^\uparrow T^{0}] & 0 & 
        \liou[D^\uparrow S]\\
        0& \liou[D^\downarrow D^\downarrow] & 0 & \liou[D^\downarrow T^{0}] & \liou[D^\downarrow T^{-1}]   & 
        \liou[D^\downarrow S]\\
        \liou[T^{+1} D\uparrow] &0&\liou[T^{+1}T^{+1}]&0&0&0\\
        \liou[T^{0} D\uparrow] &\liou[T^{0} D\uparrow]&0&\liou[T^0T^0]&0&0\\
        0& \liou[T^{-1} D\uparrow] &0 &0&\liou[T^{-1}T^{-1}]&0\\
        \liou[S D\uparrow] &\liou[S D\uparrow]&0&0&0&\liou[SS]\\
    \end{pmatrix}.
    \end{aligned}
\end{equation}
Explicitly, we have the term describing the depopulation of the doublet states towards the anionic states
\begin{equation}
    \begin{aligned}
        &\liou[D^\uparrow D^\uparrow] = \liou[D^\downarrow D^\downarrow] = \\
        &-\gsz (3f^+_{\rm sub}(\epsilon_\mathrm{t})+f^+_{\rm sub}(\epsilon_\mathrm{s}))
      \\&-\gtz(3f^+_{\rm tip}(\epsilon_\mathrm{t}) +f^+_{\rm tip}(\epsilon_\mathrm{s})),
    \end{aligned}
\end{equation}
with $\epsilon_{\mathrm{t/s}}$ introduced in \cref{eq:pseudospin_tun}.

The depopulation of the anionic states are $2\times 2$ matrices which read for the triplets
\begin{equation}
    \begin{aligned}
        &\liou[T^{+1}T^{+1}] = \liou[T^{0}T^{0}]=\liou[T^{-1}T^{-1}]=\\
        &\begin{pmatrix}
            -\gsz f^-_{\rm sub}(\epsilon_\mathrm{t})-\gtz f^-_{\rm tip}(\epsilon_\mathrm{t}) & 0 \\
            0& -\gsz f^-_{\rm sub}(\epsilon_\mathrm{t})  \\
        \end{pmatrix},
    \end{aligned}
\end{equation}
and for the singlets
\begin{equation}
    \begin{aligned}
        &\liou[SS]=\\
        &\begin{pmatrix}
            -\gsz f^-_{\rm sub}(\epsilon_\mathrm{s})-\gtz f^-_{\rm tip}(\epsilon_\mathrm{s}) & 0 \\
            0& -\gsz f^-_{\rm sub}(\epsilon_\mathrm{s})  \\
        \end{pmatrix}.
    \end{aligned}
\end{equation}
The tunneling from the $\ket{T^{-1}}$ states to $\ket{D\uparrow}$ is the same as from $\ket{T^{-1}}$ to $\ket{D^\downarrow}$
\begin{equation}
    \begin{aligned}
        &\liou[D^\uparrow T^{+1}] = \liou[D^\downarrow T^{-1}] =\\
        &\left( \gsz f^-_{\rm sub}(\epsilon_\mathrm{t})+2\gtz f^-_{\rm tip}(\epsilon_\mathrm{t}),\: \gsz f^-_{\rm sub}(\epsilon_\mathrm{t}) \right), 
    \end{aligned}
\end{equation}
whereas tunneling from the $\ket{T^0}$ states is possible to both doublet states but every channel has a smaller ''effective'' rate
\begin{equation}
    \begin{aligned}
        &\liou[D^\uparrow T^0]= \liou[D^\downarrow T^0] = \\ 
        &\left( \tfrac{\gsz}{2} f^-_{\rm sub}(\epsilon_\mathrm{t})+\gtz f^-_{\rm tip}(\epsilon_\mathrm{t}),\: \tfrac{\gsz}{2} f^-_{\rm sub}(\epsilon_\mathrm{t}) \right).
    \end{aligned}
\end{equation}
The same holds true for the singlets
\begin{equation}
    \begin{aligned}
        &\liou[D^\uparrow S]= \liou[D^\downarrow S] = \\ 
        &\left( \tfrac{\gsz}{2} f^-_{\rm sub}(\epsilon_\mathrm{s})+\gtz f^-_{\rm tip}(\epsilon_\mathrm{s}),\: \tfrac{\gsz}{2} f^-_{\rm sub}(\epsilon_\mathrm{s}) \right) .
    \end{aligned}
\end{equation}
The repopulation rates of the $\ket{T^{+1}}$ and $\ket{T^{-1}}$ states are given by
\begin{equation}
    \begin{aligned}
        &\liou[T^{+1}D^\uparrow] = \liou[T^{-1}D^\downarrow] = \\
        & \left( \gsz f^+_{\rm sub}(\epsilon_\mathrm{t})+\gtz f^+_{\rm tip}(\epsilon_\mathrm{t}),\: \gsz f^+_{\rm sub}(\epsilon_\mathrm{t}) \right)^\intercal.
    \end{aligned}
\end{equation}
Furthermore, we have the repopulation of the $\ket{T^0}$ states
\begin{equation}
    \begin{aligned}
        &\liou[T^{0}D^\uparrow] = \liou[T^{0}D^\downarrow] = \\
        & \left( \tfrac{\gsz}{2} f^+_{\rm sub}(\epsilon_\mathrm{t})+\gtz f^+_{\rm tip}(\epsilon_\mathrm{t}),\: \tfrac{\gsz}{2} f^+_{\rm sub}(\epsilon_\mathrm{t}) \right)^\intercal,
    \end{aligned}
\end{equation}
and the singlets
\begin{equation}
    \begin{aligned}
        &\liou[SD^\uparrow] = \liou[SD^\downarrow] = \\
        & \left( \tfrac{\gsz}{2} f^+_{\rm sub}(\epsilon_\mathrm{s})+\gtz f^+_{\rm tip}(\epsilon_\mathrm{s}),\: \tfrac{\gsz}{2} f^+_{\rm sub}(\epsilon_\mathrm{s}) \right)^\intercal.
    \end{aligned}
\end{equation}

By solving the eigenvalue equation
\begin{equation}
    \liou[T]\rho_i = \Gamma_i \rho_i,
\end{equation}
we associate a tunneling rate $\Gamma_i$ to an eigenstate of the tunneling Liouvillean $\rho_i$.
The $\rho_i$ form a basis and therefore we can expand the thermal density matrix in this basis as
\begin{equation}
    \rho_\mathrm{thermal} = \sum_i A_i^0 \rho_i,
\end{equation}
with expansion coefficients $A_i^0$.
The time evolution of a density matrix is governed by
\begin{equation}
\label{eq:ME}
    \dot{\rho} = \liou[T] \rho,
\end{equation}
which has the solution
\begin{equation}
    \rho(t) = e^{\liou[T] t} \rho_\mathrm{thermal} = \sum_{i}A_i^0 e^{-\Gamma_i t}\rho_i.
\end{equation}
We find for the rates $\Gamma_i$ five different energy scales
\begin{equation}
\begin{aligned}
    &\hbar \Gamma_{1/2} \sim \si{\meV},\quad \hbar\Gamma_{3-6}\sim \SI{e-1}{\meV},
    \\& \hbar \Gamma_{7} \sim \SI{e-35}{\meV}, \quad \hbar \Gamma_{8/9} \sim \SI{e-38}{\meV},
    \\& \hbar \Gamma_{10} = \SI{0}{\meV}.
\end{aligned}
\end{equation}
The eigenvectors $\rho_i$ with $i=1-6$ thus do not contribute to the density matrix on the timescale set by the pump-laser pulse. The state $\rho_{10}$ represents the steady-state solution of the master equation given by the thermal distribution of the populations of the coupled singlet and triplet states.
The three remaining eigenstates $\rho_{7-9}$ do not decay on the time-scales of our simulations.  The depopulating rate of a singlet state via a substrate tunneling transitions sets the order of magnitude of $\Gamma_7 \sim \gsz \fmsS$, whereas  $\Gamma_{8/9} \sim \gsz \fms$ are of the order of the substrate depopulating rate of a triplet state.

\section{Free evolution of the pseudospin components $\langle \tau_y\rangle$ and $\langle \tau_z\rangle$}
\label{sec:tytz_evolution}

We give here the analytical expression of the time evolution of the $y$ and $z$ components of the pseudospin solely under the influence of $\hamil[SOI]$. 
The $y$ component reads
\begin{equation}
    \begin{aligned}
        \ev{\tau_y}(t) = &\frac{1}{3}
        \big(\evz{\tau_y^\mathrm{t}}-\evz{S_{z^2}^\mathrm{t}\tau_y^\mathrm{t}}\big) +\evz{\tau_y^\mathrm{s}}+\\ 
        \frac{1}{12}\bigg[&\sqrt{a_y^2+c_y^2}\cos\left(\omega_{3} t- \arctan\frac{c_y}{a_y}\right) +\\&
        \sqrt{b_y^2+d_y^2}\cos\left(\omega_{4} t-\arctan\frac{d_y}{b_y}\right)\bigg],
    \end{aligned}
\end{equation}
where the two frequencies are given by 
\begin{equation}
\omega_{3} = \frac{1}{\hbar}(\alpha_1-\alpha_2-\alpha_3)\quad  \text{and} \quad 
\omega_{4} = \frac{1}{\hbar}(\alpha_1+\alpha_2+\alpha_3).
\end{equation}
The phases and amplitudes of the oscillations are obtained from the four parameters
\begin{equation}
    \begin{aligned}
    a_y &= 4\evz{\tau_y^{\rm t}}+3\evz{S_{xy}^{\rm t}}+2\evz{S_{z^2}^{\rm t}\tau_y^{\rm t}}, \\
    b_y &= 4\evz{\tau_y^{\rm t}}-3\evz{S_{xy}^{\rm t}}+2\evz{S_{z^2}^{\rm t}\tau_y^{\rm t}}, \\
    c_y &= 6\big(\evz{S_{z}^{\rm t}\tau_x^{\rm t}}+2\evz{S_{x^2y^2}^{\rm t}\tau_z^{\rm t}}\big), \\
    d_y &= 6\big(\evz{S_{z}^{\rm t}\tau_x^{\rm t}}-2\evz{S_{x^2y^2}^{\rm t}\tau_z^{\rm t}}\big), \\
    \end{aligned}
\end{equation} 
which can be calculated from the several expectation values and correlators calculated at the initial time $t = 0$. 
The $z$ component does only exhibit slower dynamics and is described by
\begin{equation}
        \begin{aligned}
            \ev{\tau_z} = &\frac{1}{3}\left(\evz{\tau_z^\mathrm{t}}-\evz{S_{z^2}^\mathrm{t}\tau_z^\mathrm{t}}\right)+\evz{\tau_z^\mathrm{s}}\\
        \frac{1}{12}\bigg[&
        \sqrt{a_z^2+c_z^2}\cos\left(\frac{\alpha_2}{\hbar} t- \arctan\frac{c_z}{a_z}\right) +\\&
        \sqrt{b_z^2+d_z^2}\cos\left(\frac{\alpha_3}{\hbar} t-\arctan \frac{d_z}{b_z}\right)\bigg],
    \end{aligned}
\end{equation}
with 
\begin{equation}
    \begin{aligned}
    a_z &= 4\evz{\tau_z^{\rm t}}-3\evz{S_z^{\rm t}}  +2\evz{S_{z^2}^{\rm t}\tau_z^{\rm t}}, \\
    b_z &= 4\evz{\tau_z^{\rm t}}+3\evz{S_z^{\rm t}}  +2\evz{S_{z^2}^{\rm t}\tau_z^{\rm t}}, \\
    c_z & = 3\big(\evz{S_{x^2y^2}^{\rm t}\tau_y^{\rm t}}-\evz{S_{xy}^{\rm t}\tau_x^{\rm t}}\big),\\
    d_z & = 3\big(\evz{S_{x^2y^2}^{\rm t}\tau_y^{\rm t}}+\evz{S_{xy}^{\rm t}\tau_x^{\rm t}}\big).
    \end{aligned}
\end{equation}

\bibliography{references}
\end{document}